\documentclass[11pt]{article}
\usepackage{graphicx}
\usepackage{latexsym}
\usepackage{amssymb}
\usepackage{color}
\usepackage{amsfonts}
\usepackage{amsmath}
\usepackage{mathtools}
\usepackage{booktabs}
\usepackage{url}
\usepackage{comment}
\usepackage{afterpage}
\usepackage{epstopdf}
\usepackage{graphics}
\usepackage[center]{subfigure}
\newcommand{\BA}{\begin{array}}
\newcommand{\EA}{\end{array}}

\newcommand{\CP}{\stackrel{p}{\longrightarrow}}

\newtheorem{theorem}{Theorem}

\newtheorem{remark}{Remark}
\newtheorem{assumption}{Assumption}

\newcommand{\argmax}{\operatornamewithlimits{argmax}}

\usepackage[numbers,square,sort]{natbib} 
\bibpunct{(}{)}{,}{a}{,}{,} 

\begin{document}
\baselineskip= 7mm

\title{Confidence Sets for the Emergence, Collapse, and Recovery Dates of a Bubble\footnote{\footnotesize{We thank two anonymous refreees,  professors Yuta Koike, Tatsushi Oka, Yohei Yamamoto, Daisuke Yamazaki, and the participants at the seminar at Hitotsubashi University, the K4 conference held at Kobe University, and 2025 World congress of the econometric society for helpful comments. ChatGPT5.1 was used for the language improvement. All errors are our responsibility. Kurozumi's research was supported by JSPS KAKENHI Grant Number 22K01422, 23H00804, and 25K05034. Skrobotov's research was supported by the Basic Research Program at HSE University.
 Address correspondence to Eiji Kurozumi, Department of Economics, Hitotsubashi University, 2-1 Naka, Kunitachi, Tokyo 186-8601, Japan; e-mail: kurozumi@econ.hit-u.ac.jp
}}}
\author{
Eiji Kurozumi$^{a}$, Anton Skrobotov$^{b}$ \\
{\small {$^{a}$ Hitotsubashi University}}\\
{\small {$^{b}$ Center for Big Data in Economics and Finance, HSE University}}
}
\date{\today}
\maketitle

\begin{abstract}
We propose constructing confidence sets for the emergence, collapse, and recovery dates of a bubble separately by inverting tests for the location of the break date. We examine both likelihood ratio-type tests and the Elliott-M\"uller-type (2007) tests for detecting break locations. The limiting distributions of these tests are derived under the null hypothesis, and their asymptotic consistency under the alternative is established. Finite-sample properties are evaluated through Monte Carlo simulations. The results indicate that combining different types of tests effectively controls the empirical coverage rate while maintaining a reasonably small length of the confidence set.
\medskip


\medskip


\end{abstract}

\newpage
\section{Introduction}

Since the seminal work by \citet{PhillipsWY2011a} (hereafter PWY), numerous tests for bubble detection have been developed in the literature. In PWY, the supremum of right-tailed unit root tests in subsamples is proposed to detect bubbles and shown to have considerable power compared with the full-sample test. On the other hand, the double-supremum test developed by \citet{PhillipsSY2015a, PhillipsSY2015b} (hereafter PSY) is effective for detecting multiple bubbles. These subsample right-tailed ADF tests have become standard and are widely used to identify explosive behavior in financial asset prices. The PWY and PSY tests have since been extended in several directions. For instance, the model has been adapted to account for shocks with nonstationary volatility, and robust tests have been developed by \citet{HarveyLST2016}, \citet{HarveyLZ2019}, \citet{HarveyLZ2020}, \citet{KurozumiST2023}, and \citet{HarveyLTZ2024}, among others. \citet{KurozumiNishi2024} investigated the random-coefficient model with a possible explosive root and demonstrated that some tests are more powerful than the right-tailed ADF tests. In addition, \citet{WuShiWu2024} showed that quantile regression techniques are effective for bubble detection. For a comprehensive review of bubble detection methods, see \citet{Skrobotov2023}.

Once evidence of a bubble is detected, attention often turns to estimating its emergence, collapse, and recovery dates. PWY, PSY, and \citet{PhillipsS2018}, among others, proposed date-stamping methods based on bubble detection tests and demonstrated that their estimators are consistent for the break fractions. Alternatively, traditional methods for detecting break dates, which minimize the sum of squared residuals, have been explored by \citet{HarveyLS2017}, \citet{PangCZL2018}, \citet{HarveyLW2020}, \citet{PangDC2021}, \citet{KurozumiS2023}, and \citet{KejriwalNP2025}. These studies show that the emergence, collapse, and recovery dates can be consistently estimated if the explosive speed is sufficiently fast, whereas only consistency for the break fractions is guaranteed for the emergence and recovery dates under slow explosiveness. However, to the best of our knowledge, confidence sets for the bubble dates have not yet been developed in the literature.

In this article, we address this gap by investigating methods for constructing confidence sets for the emergence, collapse, and recovery dates of a bubble. A commonly used approach for constructing a confidence set for a break date relies on the limiting distribution of the breakpoint estimator. However, our preliminary simulations reveal that this method performs poorly for bubble dates. Instead, we propose constructing the confidence set by inverting tests for the location of the break date, as proposed by \citet{ElliottM2007} and \citet{EoM2015}, among others. We develop several tests, including one based on the likelihood ratio principle, as in \citet{EoM2015}, and others inspired by \citet{ElliottM2007}. Note that we construct the confidence sets for the three breaks not jointly but separately, given the point estimates of the breaks. We examine the asymptotic properties of these tests under both the null and alternative hypotheses. Furthermore, we evaluate their finite-sample performance through Monte Carlo simulations, showing that some of the proposed methods perform satisfactorily in terms of both the coverage rate and the length of the confidence set.

The remainder of this article is organized as follows. Section~2 introduces the model and assumptions. Section~3 proposes several methods for constructing confidence sets for the emergence, collapse, and recovery dates of the bubble and develops the associated asymptotic theory. Section~4 investigates the finite-sample properties, and Section~5 presents an empirical application. Concluding remarks are given in Section~6. Proofs of the theorems are relegated to the appendix, and detailed simulation results are provided in the online appendix.

\section{Model and Assumptions}

We consider the following model for the emergence and collapse of a bubble for $t = 1, 2, \ldots, T$:
\begin{align}
    y_t &= 
    \begin{cases}
    \mu + y_{t-1} + \varepsilon_t, & 1 \leq t \leq T_e,\\
    \phi_a y_{t-1} + \varepsilon_t, & T_e + 1 \leq t \leq T_c,\\
    \phi_b y_{t-1} + \varepsilon_t, & T_c + 1 \leq t \leq T_r,\\
    \mu + y_{t-1} + \varepsilon_t, & T_r + 1 \leq t \leq T,
    \end{cases}
    \label{model:0}
\end{align}
where $y_0 = o_p(T^{1/2})$, $\phi_a \coloneqq 1 + a/T^{\alpha}$ with $a > 0$ and $0 < \alpha < 1$, and $\phi_b \coloneqq 1 - b/T^{\beta}$ with $b > 0$ and $0 < \beta < 1$. The drift term $\mu$ is sometimes specified as being local to zero, such as $\mu = c/T^{\gamma}$ with $1/2 < \gamma \leq 1$, in which case the effect of the drift term becomes negligible. For simplicity, we assume $\mu = 0$ throughout this article.

In model~\eqref{model:0}, the process $\{y_t\}$ follows a random walk in the first regime but becomes explosive at $t = T_e + 1$. It then collapses at $t = T_c + 1$, where $\{y_t\}$ becomes mean-reverting with $\phi_b < 1$, and eventually returns to a random walk. The emergence, collapse, and recovery dates ($T_e$, $T_c$, and $T_r$, respectively) can thus be viewed as the break dates of the AR(1) coefficients. The corresponding break fractions are defined as $\lambda_e \coloneqq T_e/T$, $\lambda_c \coloneqq T_c/T$, and $\lambda_r \coloneqq T_r/T$. We assume that these fractions are distinct, satisfying $0 < \lambda_e < \lambda_c < \lambda_r < 1$, as is standard in the literature.

For model~\eqref{model:0}, we impose the following assumption.

\begin{assumption}
\label{assumption:error}
The sequence $\{\varepsilon_t\}$ for $t = 1, 2, \ldots$ is a martingale difference sequence with $E[\varepsilon_t | \mathcal{F}_{t-1}] = 0$ a.s.\ for all $t$ and $\sup_t E[\varepsilon_t^2] < \infty$, where $\mathcal{F}_t = \sigma\{\varepsilon_t, \varepsilon_{t-1}, \ldots\}$. We assume that the following law of large numbers and functional central limit theorem (FCLT) hold:
\begin{equation}
    \frac{1}{T}\sum_{t=1}^{\lfloor Tr \rfloor} E[\varepsilon_t^2] \to r\sigma^2, 
    \quad 
    \frac{1}{T}\sum_{t=1}^{\lfloor Tr \rfloor} \varepsilon_t^2 \xrightarrow{p} r\sigma^2,
    \label{WLLN}
\end{equation}
for $0 \leq r \leq 1$, where $\xrightarrow{p}$ denotes convergence in probability, and
\begin{equation}
    \frac{1}{\sqrt{T}}\sum_{t=1}^{\lfloor Tr \rfloor} \varepsilon_t \Rightarrow \sigma W(r),
    \label{FCLT}
\end{equation}
where $W(\cdot)$ is a standard Brownian motion and $\Rightarrow$ denotes weak convergence of the associated probability measures.
\end{assumption}

Assumption~\ref{assumption:error} requires the shocks to be uncorrelated but allows for conditional heteroskedasticity. Note that the pointwise convergences in \eqref{WLLN} imply uniform convergence over $0 \leq r \leq 1$ because the functions are monotonic in $r$ (see, for example, Lemma~A.10 in \cite{Hansen2000b}). The FCLT in \eqref{FCLT} holds, for example, when $\{\varepsilon_t\}$ is strongly uniformly integrable under certain additional conditions (see Section~3.6 in \cite{TANAKA96}).

\section{Construction of the Confidence Sets}

In this section, we consider several methods for constructing confidence sets for $T_e$, $T_c$, and $T_r$, and investigate their asymptotic properties. Several approaches for estimating the bubble dates have been proposed in the literature, and all of them establish the consistency of the break fraction estimators. Hence, regardless of the specific estimation method employed, we focus on subsamples divided according to the estimated break dates. In particular, we use the estimators $\hat{T}_e$, $\hat{T}_c$, and $\hat{T}_r$ obtained via the step-by-step least squares method proposed by \citet{KurozumiS2023}. We also employ the OLS estimators $\hat{\phi}_a$ and $\hat{\phi}_b$ obtained from the corresponding subsamples, while $\hat{\sigma}^2$ is computed from the residuals of the full sample. It is straightforward to show that $\hat{\phi}_a / \phi_a \xrightarrow{p} 1$, $\hat{\phi}_b / \phi_b \xrightarrow{p} 1$, and $\hat{\sigma}^2 \xrightarrow{p} \sigma^2$.

In the following subsections, since the break fraction estimators are consistent, as shown in \citet{KurozumiS2023}, we proceed with the analysis using the true regimes for ease of exposition: $[1, T_c]$, $[T_e + 1, T_r]$, and $[T_c + 1, T]$, corresponding to the analyses of the emergence date $T_e$, the collapse date $T_c$, and the recovery date $T_r$, respectively.

\subsection{Confidence Set for the Emergence Date of a Bubble}

In this subsection, we focus on the emergence date of a bubble in the regime $[1, T_c]$ and construct a confidence set for the emergence date using the first two regimes, in which $T_e$ is unknown:
\begin{align}
    y_t &=
    \begin{cases}
    y_{t-1} + \varepsilon_t, & 1 \leq t \leq T_e,\\
    \phi_a y_{t-1} + \varepsilon_t, & T_e + 1 \leq t \leq T_{c}.
    \end{cases}
    \label{model:emerging}
\end{align}
Let $T_{UB} \coloneqq T_c$ denote the sample size of the first two regimes.

\subsubsection{Confidence Interval Based on the Asymptotic Distribution of the Breakpoint Estimator}

One frequently used method for constructing a confidence set for a break date is based on the limiting distribution of the breakpoint estimator. For the emergence date of a bubble, \citet{PangCZL2018} show in their Theorem~1.3 that
\begin{equation}
    T_{UB}^2 \rho_a^2 \bigl(\hat{\lambda}_e^* - \lambda_e^*\bigr) \Rightarrow \argmax_{v \in \mathbb{R}}
    \left\{\frac{W^*(v)}{W_1(\lambda_e^*)} - \frac{|v|}{2}\right\}
    \label{lambda_e:limit_dist}
\end{equation}
when $\alpha > 1/2$, where $\hat{\lambda}_e^* \coloneqq \hat{T}_e/T_{UB}$ is the estimator of $\lambda_e^* \coloneqq T_e/T_{UB}$ (the relative position of $T_e$ within the first two regimes), obtained by minimizing the sum of squared residuals in \eqref{model:emerging}; $\rho_a \coloneqq \phi_a - 1$; and $W^*(\cdot)$ is a two-sided Brownian motion on $\mathbb{R}$ defined by $W^*(v) \coloneqq W_1(-v)$ for $v \leq 0$ and $W^*(v) \coloneqq W_2(v)$ for $v > 0$. For $\alpha = 1/2$, the breakpoint estimator is $O_p(1)$, whereas $\hat{T}_e$ is consistent for $T_e$ when $\alpha < 1/2$. In other words, the emergence date can be consistently estimated if the explosive speed is sufficiently fast.

Although the asymptotic behavior of $\hat{T}_e$ depends on the true value of $\alpha$, we may construct a confidence interval for $T_e$ based on \eqref{lambda_e:limit_dist} under the assumption $\alpha > 1/2$, namely
\[
\left[\hat{T}_e - \left\lceil \frac{c_u}{\hat{\rho}_a^2 T} \right\rceil,\;
      \hat{T}_e - \left\lfloor \frac{c_l}{\hat{\rho}_a^2 T} \right\rfloor \right],
\]
where $\hat{\rho}_a \coloneqq \hat{\phi}_a - 1$ and $c_l$ and $c_u$ are the 0.05 and 0.95 quantiles of the limiting distribution in \eqref{lambda_e:limit_dist} for a 0.9 confidence level. However, our preliminary simulations indicate that this interval tends to be extremely wide: the empirical coverage rate is essentially one and the interval covers all permissible break dates. Because of this poor performance, we do not further pursue confidence intervals based on the limiting distribution of the breakpoint estimator.

\subsubsection{Confidence Set Based on the LR Test}

Another approach is to invert a test for the break date. Consider
\[
H_0:\; T_e = T_1 \qquad \text{vs.} \qquad H_1:\; T_e=T_2 \neq T_1,
\]
and exclude $T_1$ from the confidence set if the null is rejected; otherwise, include it. By repeatedly conducting the test over the permissible range of $T_1$ at significance level $\delta$, we obtain a confidence set with nominal confidence $1-\delta$. In this framework, the correctness of the coverage rate is tied to the empirical size of the test, while the length of the confidence set depends on its power.

For the test construction, assume $\{\varepsilon_t\} \sim i.i.d.\ N(0, \sigma^2)$.\footnote{As pointed out by one of the referees, tests for structural change may be used to construct confidence sets for the break date if they have sufficient discriminatory power. However, it is more natural to use tests designed for hypotheses about the location of the break date, and we therefore focus on such tests in this article.} This assumption is used only to derive the test statistic; the asymptotic theory below does not rely on normality. A natural candidate is a likelihood ratio (LR)-type test, as in \citet{EoM2015} for regime-wise stationary models. Pin down the alternative to $H_1:\; T_e = T_2$ ($T_2 \neq T_1$) and let the joint density of $\mathbf{y} \coloneqq \{y_1, \ldots, y_T\}$ be $f_T(\mathbf{y}; H_i)$ under $H_i$. The LR principle suggests rejecting $H_0$ when $f_T(\mathbf{y}\mid H_1)/f_T(\mathbf{y}\mid H_0)$ is large. It can be shown that an equivalent rejection rule is
\begin{align}
    2\sum_{t=T_1+1}^{T_2} y_{t-1}\Delta y_t - \rho_a \sum_{t=T_1+1}^{T_2} y_{t-1}^2  &< c_1 \quad \text{for } T_1 < T_2,
    \label{LR12:org}\\
    2\sum_{t=T_2+1}^{T_1} y_{t-1}\Delta y_t - \rho_a \sum_{t=T_2+1}^{T_1} y_{t-1}^2  &> c_2 \quad \text{for } T_2 < T_1,
    \label{LR21:org}
\end{align}
where $c_1$ and $c_2$ are chosen by the significance level. Under $H_0: T_e = T_1$, the convergence rates of \eqref{LR12:org} and \eqref{LR21:org} differ, so distinct scalings are needed depending on whether $T_1 < T_2$ or $T_2 < T_1$. We therefore propose, for $T_1 < T_2$,
\begin{equation}
    \mathrm{LR}_{a,12}^e
    \coloneqq
    \frac{\displaystyle \min_{T_2 \in \Lambda_{12}^e} \left( 2\sum_{t=T_1+1}^{T_2} y_{t-1}\Delta y_t - \hat{\rho}_a \sum_{t=T_1+1}^{T_2} y_{t-1}^2 \right)}
         {\displaystyle T \hat{\phi}_a^{2(T_{LRa12}^e - T_1)} \hat{\sigma}^2/2},
    \label{LRa12}
\end{equation}
where $\Lambda_{12}^e \coloneqq \{ T_2: T_1 + \lfloor T_{UB}\epsilon \rfloor \le T_2 \le T_c \}$ is the permissible range of $T_2$ with trimming parameter $\epsilon$, and $T_{LRa12}^e$ is the minimizer of the numerator of \eqref{LRa12}. For $T_2 < T_1$, define
\begin{equation}
    \mathrm{LR}_{a,21}^e
    \coloneqq
    \frac{\displaystyle \max_{T_2 \in \Lambda_{21}^e} \left( 2\sum_{t=T_2+1}^{T_1} y_{t-1}\Delta y_t - \hat{\rho}_a \sum_{t=T_2+1}^{T_1} y_{t-1}^2 \right)}
         {\displaystyle T_{UB}^2 \hat{\rho}_a \hat{\sigma}^2},
    \label{LRa21}
\end{equation}
where $\Lambda_{21}^e \coloneqq \{T_2: 1 \le T_2 \le T_1 - \lfloor T_{UB}\epsilon \rfloor\}$. Because we do not know whether $T_1 < T_e$ or $T_1 > T_e$ under the alternative, we combine $\mathrm{LR}_{a,12}^e$ and $\mathrm{LR}_{a,21}^e$ and reject $H_0: T_e = T_1$ if
\begin{equation}
\mathrm{LR}_{a,12}^e < \mathrm{cv}_{LR12,\delta/2}^e
\quad \text{or} \quad
\mathrm{LR}_{a,21}^e > \mathrm{cv}_{LR21,1-\delta/2}^e,
\label{LRatest}
\end{equation}
where $\mathrm{cv}_{LR12,\delta/2}^e$ and $\mathrm{cv}_{LR21,1-\delta/2}^e$ are the $\delta/2$ and $1-\delta/2$ quantiles of the null limiting distributions of \eqref{LRa12} and \eqref{LRa21}, respectively. We call this the $\mathrm{LR}_a^e$ test. Note that
\[
P\!\left(\mathrm{LR}_{a,12}^e < \mathrm{cv}_{LR12,\delta/2}^e \;\text{or}\; \mathrm{LR}_{a,21}^e > \mathrm{cv}_{LR21,1-\delta/2}^e\right)
\le \delta,
\]
so the asymptotic coverage is conservative (at least $1-\delta$).

\begin{remark}
A joint test based on $\mathrm{LR}_{a,12}^e$ and $\mathrm{LR}_{a,21}^e$ can control the asymptotic size via the union-of-rejections rule: reject if
\[
\frac{\mathrm{LR}_{a,12}^e}{\mathrm{cv}_{LR12,\delta/2}^e} < \psi_{LRa,\delta}^e
\quad \text{or} \quad
\frac{\mathrm{LR}_{a,21}^e}{\mathrm{cv}_{LR21,1-\delta/2}^e} < \psi_{LRa,\delta}^e,
\]
where the inequality for $\mathrm{LR}_{a,21}^e$ is reversed because $\mathrm{cv}_{LR21,1-\delta/2}^e$ is negative (see Theorem~\ref{theorem:emergence:H0}). Our preliminary simulations, however, indicate that the resulting confidence set is nearly identical to that from the conservative rule \eqref{LRatest} for all tests considered. Since the union rule requires the additional critical value $\psi_{LRa,\delta}^e$ (in addition to two critical values), we focus on the conservative procedure \eqref{LRatest}.
\end{remark}

As shown in the simulation section, the above LR tests---especially $\mathrm{LR}_{a,12}^e$---tend to be over-sized in finite samples. Because the rejection region is in the left tail, over-sizing indicates that the statistic is too small in finite samples. To mitigate this distortion, we propose a finite-sample correction that shifts the statistic in the positive direction without altering its asymptotic null distribution. Noting that
\[
2\sum_{t=T_1+1}^{T_2} y_{t-1}\Delta y_t
=
y_{T_2}^2 - y_{T_1}^2 - \sum_{t=T_1+1}^{T_2} (\Delta y_t)^2 \le y_{T_2}^2,
\]
and that only $y_{T_2}^2$ contributes asymptotically in the above expansion, we modify $\mathrm{LR}_{a,12}^e$ to
\begin{equation}
    \mathrm{LR}_{b,12}^e
    \coloneqq
    \frac{\displaystyle \min_{T_2 \in \Lambda_{12}^e} \left( y_{T_2}^2 - \hat{\rho}_a \sum_{t=T_1+1}^{T_2} y_{t-1}^2 \right)}
         {\displaystyle T \hat{\phi}_a^{2(T_{LRb12}^e - T_1)} \hat{\sigma}^2 / 2},
    \label{LRb12}
\end{equation}
where $T_{LRb12}^e$ minimizes the numerator of \eqref{LRb12}. We will combine $\mathrm{LR}_{b,12}^e$ with $\mathrm{EM}_{a,21}^e$, introduced next.

\subsubsection{Confidence Set Based on the EM-Type Test}

Optimal tests for break dates in regime-wise stationary regression models have been studied by \citet{ElliottM2007}, \citet{KurozumiYamamoto2015}, and \citet{Yamamoto2018}, among others, and for cointegrating regressions by \citet{KurozumiSkrobotov2018}. These works assume priors on the break size and break date and derive tests that maximize weighted average power. Typically, the prior for the break size is normal, chosen so that the test is invariant to the magnitude of the break, while the break-date prior is uniform. As a result, the optimal test can be written as a weighted average of LR statistics:
\begin{equation}
    \int \int \frac{f_T(\mathbf{y}; H_1)}{f_T(\mathbf{y}; H_0)} \, dQ_{T_2}(a)\, dJ(T_2),
    \label{optimaltest}
\end{equation}
where $Q_{T_2}(a)$ and $J(T_2)$ are the priors. Although the formal optimality of \eqref{optimaltest} requires exogenous regressors (not our case), it nonetheless suggests suitable candidate statistics for testing the break date.

For tractability, suppose $\{\varepsilon_t\} \sim i.i.d.\ N(0,\sigma^2)$ and let
\[
Q_{T_2}(a) \sim N\!\left(0,\; \kappa\left(\frac{1}{\sigma^2 T^{2\alpha}} \sum_{t=T_2+1}^{T_1} y_{t-1}^2 \right)^{-1}\right)
\]
when $T_1 > T_2$, where $\kappa > 0$ controls the dispersion of $a$. This prior is adopted solely to obtain a tractable statistic. Conditional on $\{y_t\}$, the integral over $a$ in \eqref{optimaltest} is explicit for $\kappa < 2$; direct calculation yields
\[
    \int \int \frac{f_T(\mathbf{y}; H_1)}{f_T(\mathbf{y}; H_0)} \, dQ_{T_2}(a)\, dJ(T_2)
    \propto
    \int \left(\frac{2 - \kappa}{2}\right)^{-1/2}
    \exp\!\left[
      \frac{\kappa}{2(2-\kappa)}
      \frac{\bigl(\sum_{t=T_1+1}^{T_2} y_{t-1}\Delta y_t\bigr)^2}
           {\sigma^2 \sum_{t=T_1+1}^{T_2} y_{t-1}^2}
    \right] dJ(T_2).
\]
This suggests basing the test on
\[
t_{T_1,T_2}^2 \coloneqq 
\frac{\displaystyle \left(\sum_{t=T_1+1}^{T_2} y_{t-1}\Delta y_t\right)^2}
     {\displaystyle \hat{\sigma}^2 \sum_{t=T_1+1}^{T_2} y_{t-1}^2},
\]
the square of the $t$-statistic for $\rho$ in a regression of $\Delta y_t$ on $y_{t-1}$ over $t=T_1+1,\ldots,T_2$, with $\hat{\sigma}^2$ from the full-sample regression as before. Analogously, $t_{T_2,T_1}^2$ serves against the alternative $T_1 > T_2$. Following \citet{Yamazaki2021}, using one-sided $t$-statistics can shrink the confidence set when the direction of the break is known; since the AR coefficient moves from a unit root to an explosive value here, we employ one-sided statistics.

We consider average-type statistics with suitable scalings:
\begin{equation}
\mathrm{EM}_{a,12}^e \coloneqq 
\frac{\displaystyle \sum_{T_2 = T_1+\lfloor T_{UB}\epsilon \rfloor}^{T_c} t_{T_1,T_2}}
     {\sqrt{T \hat{\phi}_a^{2(T_c - T_1)} /(2\hat{\rho}_a)}}
\quad \text{and} \quad
\mathrm{EM}_{a,21}^e \coloneqq 
\frac{1}{T_{UB}}\sum_{T_2 = 1}^{T_1 - \lfloor T_{UB}\epsilon \rfloor} t_{T_2,T_1},
\label{EMa}
\end{equation}
and reject when
\[
    \mathrm{EM}_{a,12}^e < \mathrm{cv}_{EM12,\delta/2}^e
    \quad \text{or} \quad
    \mathrm{EM}_{a,21}^e > \mathrm{cv}_{EMa21,1-\delta/2}^e,
\]
where $\mathrm{cv}_{EM12,\delta/2}^e$ and $\mathrm{cv}_{EMa21,1-\delta/2}^e$ are the $\delta/2$ and $1-\delta/2$ quantiles of the respective null limiting distributions. We call this the $\mathrm{EM}_a^e$ test.

We also consider minimum- and maximum-type statistics:
\begin{equation}
    \mathrm{EM}_{b,12}^e \coloneqq 
    \frac{\displaystyle \min_{T_2 \in \Lambda_{12}^e} t_{T_1,T_2}}
         {\displaystyle \sqrt{T \hat{\rho}_a \hat{\phi}_a^{2(T_{EMb12}^e - T_1)} / 2}}
    \quad \text{and} \quad
    \mathrm{EM}_{b,21}^e \coloneqq \max_{T_2 \in \Lambda_{21}^e} t_{T_2,T_1},
    \label{EMb}
\end{equation}
where $T_{EMb12}^e$ minimizes the numerator of $\mathrm{EM}_{b,12}^e$. Reject when
\[
    \mathrm{EM}_{b,12}^e < \mathrm{cv}_{EMb12,\delta/2}^e
    \quad \text{or} \quad
    \mathrm{EM}_{b,21}^e > \mathrm{cv}_{EMb21,1-\delta/2}^e,
\]
with critical values from the corresponding null limiting distributions. Since the null limiting distributions of $\mathrm{EM}_{b,12}^e$ and $\mathrm{EM}_{a,12}^e$ coincide, we use the same critical values for the tests. Taking finite-sample performance into account (see the next section), we propose combining $\mathrm{LR}_{b,12}^e$ and $\mathrm{EM}_{a,21}^e$, which we call the $\mathrm{LE}^e$ test.

\subsubsection{Limiting Distributions of the Tests for the Emergence Date}

We first establish the limiting distributions of the test statistics under the null.

\begin{theorem}
    \label{theorem:emergence:H0}
    Suppose Assumption~\ref{assumption:error} holds. Under $H_0:\, T_e = T_1$,
    \begin{align}
        \mathrm{LR}_{a,12}^e,\; \mathrm{LR}_{b,12}^e
        & \Rightarrow
        W^2(\lambda_1),
        \label{limit:LR12:H0} \\
        \mathrm{EM}_{a,12}^e,\; \mathrm{EM}_{b,12}^e
        & \Rightarrow 
        |W(\lambda_1)|,
        \label{limit:EM12:H0}
    \end{align}
    and
    \begin{align}
        \mathrm{LR}_{a,21}^e
        & \Rightarrow 
        - \int_{\lambda_1^* - \epsilon}^{\lambda_1^*} W^2(s^*)\, ds^*,
        \label{limit:LR21:H0}\\
        \mathrm{EM}_{a,21}^e
        & \Rightarrow 
        \int_{0}^{\lambda_1^* - \epsilon} ADF(\lambda_2^*, \lambda_1^*)\, d\lambda_2^*,
        \label{limit:EMa21:H0} \\
        \mathrm{EM}_{b,21}^e
        & \Rightarrow 
        \sup_{0 \le \lambda_2^* \le \lambda_1^* - \epsilon} ADF(\lambda_2^*, \lambda_1^*),
        \label{limit:EMbc21:H0}
    \end{align}
    where $\lambda_j^* \coloneqq T_j/T_{UB}$ for $j=1,2$, and
    \[
    ADF(\lambda_2^*, \lambda_1^*) \coloneqq 
    \frac{\frac{1}{2}\bigl[ W^2(\lambda_1^*) - W^2(\lambda_2^*) - (\lambda_1^* - \lambda_2^*) \bigr]}
         {\sqrt{\int_{\lambda_2^*}^{\lambda_1^*} W^2(s^*)\, ds^*}}.
    \]
\end{theorem}

The left-tail critical values of \eqref{limit:LR12:H0} and \eqref{limit:EM12:H0} for a given $\lambda_1$ follow from chi-square percentiles:
\[
\mathrm{cv}_{LR12,0.05}^e = \lambda_1 \chi_{1,0.05}^2 = 0.00393\, \lambda_1,
\qquad
\mathrm{cv}_{EM12,0.05}^e = \sqrt{\lambda_1 \chi_{1,0.05}^2} = \sqrt{0.00393\, \lambda_1},
\]
at significance level $0.05$ (corresponding to confidence level $0.9$), where $\chi_{1,0.05}^2$ is the $0.05$ quantile of the chi-square distribution with one degree of freedom. In contrast, the critical values for \eqref{limit:LR21:H0}--\eqref{limit:EMbc21:H0} are nonstandard. We therefore compute them for $\lambda_1^* = 0.10, 0.11, \ldots, 0.90$ with $\epsilon = 0.1$ via simulation with 50{,}000 replications, approximating standard Brownian motion by scaled partial sums of 1{,}000 i.i.d.\ standard normals. As it is inconvenient to tabulate critical values for every admissible break date, we fit the response-surface regression
\[
\mathrm{cv}_{\ell 21,0.95}^e
= a_{0,\ell} + a_{-1,\ell}\frac{1}{\lambda_1^*}
  + a_{1,\ell}\lambda_1^* + a_{2,\ell}\lambda_1^{*2} + a_{3,\ell}\lambda_1^{*3},
\]
for $\ell \in \{LR, EMa, EMb\}$. The estimated coefficients are summarized in Table~1.

Next, we study the asymptotic behavior under the alternative.

\begin{theorem}
    \label{theorem:emergence:H1}
    Suppose Assumption~\ref{assumption:error} holds.
    \begin{itemize}
    \item[(i)] If $T_e > T_1$ ($\lambda_e > \lambda_1$), then
    \[
    \mathrm{LR}_{a,12}^e,\; \mathrm{LR}_{b,12}^e,\; \mathrm{EM}_{a,12}^e,\; \mathrm{EM}_{b,12}^e \xrightarrow{p} 0,
    \quad \text{whereas} \quad
    \mathrm{LR}_{a,21}^e,\; \mathrm{EM}_{a,21}^e,\; \mathrm{EM}_{b,21}^e = O_p(1).
    \]
    \item[(ii)] If $T_e < T_1$ ($\lambda_e < \lambda_1$), then
    \[
    \mathrm{LR}_{a,12}^e,\; \mathrm{LR}_{b,12}^e,\; \mathrm{EM}_{a,12}^e,\; \mathrm{EM}_{b,12}^e \to \infty,
    \quad \text{and} \quad
    \mathrm{LR}_{a,21}^e,\; \mathrm{EM}_{a,21}^e,\; \mathrm{EM}_{b,21}^e \to \infty.
    \]
    \end{itemize}
\end{theorem}

From Theorem~\ref{theorem:emergence:H1}(i), the $\mathrm{LR}_{\cdot,12}^e$ and $\mathrm{EM}_{\cdot,12}^e$ tests are consistent against $T_e > T_1$ (their rejection regions are left tails with positive critical values), whereas the $\mathrm{LR}_{a,21}^e$ and $\mathrm{EM}_{\cdot,21}^e$ tests are not. Conversely, Theorem~\ref{theorem:emergence:H1}(ii) implies that the former are inconsistent against $T_e < T_1$, while the latter are consistent. Thus, each test is consistent in exactly one direction (either $T_e > T_1$ or $T_e < T_1$), as intended in their construction.

Since we test across all permissible break dates, Theorem~\ref{theorem:emergence:H1} suggests that confidence sets based on $\mathrm{LR}_{\cdot,12}^e$ and $\mathrm{EM}_{\cdot,12}^e$ will include only a small portion of dates earlier than the true $T_e$, but may include a substantial set of dates later than $T_e$. In contrast, the $\mathrm{LR}_{a,21}^e$ and $\mathrm{EM}_{\cdot,21}^e$ tests tend to yield asymmetric confidence sets with many dates to the left of $T_e$ and relatively few to the right. These patterns are also borne out in finite samples in the next section.

\subsection{Confidence Set for the Collapse Date of a Bubble}

In this subsection, we focus on the collapse date of a bubble in the regime $[T_e+1, T_r]$ and consider a one-break model:
\begin{align}
    y_t &=
    \begin{cases}
    \phi_a y_{t-1} + \varepsilon_t, & T_e + 1 \le t \le T_c,\\
    \phi_b y_{t-1} + \varepsilon_t, & T_c + 1 \le t \le T_r.
    \end{cases}
    \label{model:collapsing}
\end{align}
Let $T_{BC} \coloneqq T_r - T_e$ denote the sample size of these two regimes.

\subsubsection{Confidence Sets for the Collapse Date}

As shown by \citet{PangDC2021} and \citet{KurozumiS2023}, the least squares estimator of the collapse date is consistent, so the limiting distribution of the breakpoint estimator is not directly useful for inference. As in the emergence-date case, we construct a confidence set by inverting a test for
\[
H_0:\; T_c = T_1 \qquad \text{vs.} \qquad H_1:\; T_c = T_2 \ne T_1.
\]
Assuming $\{\varepsilon_t\} \sim i.i.d.\ N(0, \sigma^2)$, the likelihood principle implies rejection of $H_0$ when
\begin{align}
    2\sum_{t=T_1+1}^{T_2} y_{t-1}(y_t - \phi_a y_{t-1})
    + (\phi_a - \phi_b)\sum_{t=T_1+1}^{T_2} y_{t-1}^2 &> c_1 
    \quad \text{for } T_1 < T_2,
    \label{LR12:org:c}\\
    2\sum_{t=T_2+1}^{T_1} y_{t-1}(y_t - \phi_a y_{t-1})
    + (\phi_a - \phi_b)\sum_{t=T_2+1}^{T_1} y_{t-1}^2 &< c_2 
    \quad \text{for } T_2 < T_1,
    \label{LR21:org:c}
\end{align}
which can be rewritten as
\begin{align*}
    2\sum_{t=T_1+1}^{T_2} y_{t-1}\Delta y_t + (2 - \phi_a - \phi_b)\sum_{t=T_1+1}^{T_2} y_{t-1}^2 &> c_1 
    \quad \text{for } T_1 < T_2, \\
    2\sum_{t=T_2+1}^{T_1} y_{t-1}\Delta y_t + (2 - \phi_a - \phi_b)\sum_{t=T_2+1}^{T_1} y_{t-1}^2 &< c_2 
    \quad \text{for } T_2 < T_1.
\end{align*}
Replacing unknown parameters with consistent estimators and applying appropriate scalings, we propose, for $T_1 < T_2$,
\begin{equation}
    \mathrm{LR}_{a,12}^c
    \coloneqq
    \frac{\displaystyle \max_{T_2 \in \Lambda_{12}^c} \left(
      2\sum_{t=T_1+1}^{T_2} y_{t-1}\Delta y_t
      + (2 - \hat{\phi}_a - \hat{\phi}_b)\sum_{t=T_1+1}^{T_2} y_{t-1}^2
    \right)}
    {\displaystyle T (\hat{\phi}_a - \hat{\phi}_b) \hat{\phi}_a^{\,2(T_1 - T_e)} \hat{\sigma}^2 /(2\hat{\rho}_b)},
    \label{LRa12:c}
\end{equation}
where $\hat{\rho}_b \coloneqq 1 - \hat{\phi}_b$ and $\Lambda_{12}^c \coloneqq \{T_2:\; T_1 + \lfloor T_{BC}\epsilon \rfloor \le T_2 \le T_r\}$. For $T_2 < T_1$, define
\begin{equation}
    \mathrm{LR}_{a,21}^c
    \coloneqq
    \frac{\displaystyle \min_{T_2 \in \Lambda_{21}^c} \left(
      2\sum_{t=T_2+1}^{T_1} y_{t-1}\Delta y_t
      + (2 - \hat{\phi}_a - \hat{\phi}_b)\sum_{t=T_2+1}^{T_1} y_{t-1}^2
    \right)}
    {\displaystyle T (\hat{\phi}_a - \hat{\phi}_b) \hat{\phi}_a^{\,2(T_1 - T_e)} \hat{\sigma}^2 /(2\hat{\rho}_a)},
    \label{LRa21:c}
\end{equation}
where $\hat{\rho}_a \coloneqq \hat{\phi}_a - 1$ and $\Lambda_{21}^c \coloneqq \{T_2:\; 1 \le T_2 \le T_1 - \lfloor T_{BC}\epsilon \rfloor\}$. We reject $H_0:\; T_c = T_1$ if
\begin{equation*}
\mathrm{LR}_{a,12}^c > \mathrm{cv}_{LR12,1-\delta/2}^c
\quad \text{or} \quad
\mathrm{LR}_{a,21}^c < \mathrm{cv}_{LR21,\delta/2}^c,
\label{LRatest:c}
\end{equation*}
where $\mathrm{cv}_{LR12,1-\delta/2}^c$ and $\mathrm{cv}_{LR21,\delta/2}^c$ are the $1-\delta/2$ and $\delta/2$ quantiles of the respective null limiting distributions. We refer to this as the $\mathrm{LR}_a^c$ test.

A finite-sample correction analogous to $\mathrm{LR}_{b,12}^e$ can be considered for $\mathrm{LR}_{a,12}^c$. However, preliminary simulations indicate no meaningful improvement in coverage or set length, so we do not pursue it further.

As in the emergence-date case, we may also consider EM-type tests based on the weighted average of likelihood ratios. Proceeding as before, the EM-type tests can be based on $t_{T_1,T_2}^2$ and $t_{T_2,T_1}^2$, or on one-sided $t_{T_1,T_2}$ and $t_{T_2,T_1}$ when the break direction is known. With suitable scalings, we propose
\begin{equation}
\mathrm{EM}_{a,12}^c \coloneqq 
\frac{\displaystyle \frac{1}{T_r - T_1 - \lfloor T_{BC}\epsilon \rfloor + 1}
      \sum_{T_2 = T_1 + \lfloor T_{BC}\epsilon \rfloor}^{T_r} t_{T_1,T_2}}
     {\sqrt{T \hat{\rho}_b \hat{\phi}_a^{\,2(T_1 - T_e)} / 2}},
\label{EMa12:c}
\end{equation}
and
\begin{equation}
\mathrm{EM}_{a,21}^c \coloneqq 
\frac{\displaystyle \frac{1}{T_1 - \lfloor T_{BC}\epsilon \rfloor - T_e}
      \sum_{T_2 = T_e + 1}^{T_1 - \lfloor T_{BC}\epsilon \rfloor} t_{T_2,T_1}}
     {\sqrt{T \hat{\rho}_a \hat{\phi}_a^{\,2(T_1 - T_e)} / 2}},
\label{EMa21:c}
\end{equation}
and reject when
\[
    \mathrm{EM}_{a,12}^c > \mathrm{cv}_{EM12,1-\delta/2}^c
    \quad \text{or} \quad
    \mathrm{EM}_{a,21}^c < \mathrm{cv}_{EM21,\delta/2}^c,
\]
where $\mathrm{cv}_{EM12,1-\delta/2}^c$ and $\mathrm{cv}_{EM21,\delta/2}^c$ are the respective critical values. We call this the $\mathrm{EM}_a^c$ test.

We further consider minimum- and maximum-type tests:
\begin{equation}
    \mathrm{EM}_{b,12}^c \coloneqq 
    \frac{\displaystyle \max_{T_2 \in \Lambda_{12}^c} t_{T_1,T_2}}
         {\displaystyle \sqrt{T \hat{\rho}_b \hat{\phi}_a^{\,2(T_1 - T_e)} / 2}}
    \quad \text{and} \quad
    \mathrm{EM}_{b,21}^c \coloneqq 
    \frac{\displaystyle \min_{T_2 \in \Lambda_{21}^c} t_{T_2,T_1}}
         {\sqrt{T \hat{\rho}_a \hat{\phi}_a^{\,2(T_1 - T_e)} / 2}},
    \label{EMb:c}
\end{equation}
and reject when
\[
    \mathrm{EM}_{b,12}^c > \mathrm{cv}_{EM12,1-\delta/2}^c
    \quad \text{or} \quad
    \mathrm{EM}_{b,21}^c < \mathrm{cv}_{EM21,\delta/2}^c,
\]
where the critical values coincide with those for $\mathrm{EM}_{a,12}^c$ and $\mathrm{EM}_{a,21}^c$ (see also Theorem~\ref{theorem:collapse:H0}). We refer to this as the $\mathrm{EM}_b^c$ test.

We can also consider the combination of these tests. Taking finite-sample performance into account, we propose combining $\mbox{LR}_{a,21}^c$ and $\mbox{EM}_{a,12}^c$, which we call the $\mbox{LE}^c$ test.

\subsubsection{Limiting Distributions of the Tests for the Collapse Date}

We first establish the limiting distributions under the null.

\begin{theorem}
    \label{theorem:collapse:H0}
    Suppose Assumption~\ref{assumption:error} holds. Under $H_0:\, T_c = T_1$,
    \begin{align}
    \mathrm{LR}_{a,12}^c &\Rightarrow - W^2(\lambda_e),
    \label{limit:LR12:H0:collapse} \\
    \mathrm{EM}_{a,12}^c,\; \mathrm{EM}_{b,12}^c &\Rightarrow - |W(\lambda_e)|,
    \label{limit:EM12:H0:collapse} \\
    \mathrm{LR}_{a,21}^c &\Rightarrow W^2(\lambda_e),
    \label{limit:LR21:H0:collapse} \\
    \mathrm{EM}_{a,21}^c,\; \mathrm{EM}_{b,21}^c &\Rightarrow |W(\lambda_e)|.
    \label{limit:21:H0:collapse}
    \end{align}
\end{theorem}

For a given value of $\lambda_e$, the critical values follow from chi-square percentiles:
\[
\mathrm{cv}_{LR12,0.95}^c = - \lambda_e \chi_{1,0.05}^2 = -0.00393\, \lambda_e,
\qquad
\mathrm{cv}_{LR21,0.05}^c = \lambda_e \chi_{1,0.05}^2 = 0.00393\, \lambda_e,
\]
\[
\mathrm{cv}_{EM12,0.95}^c = - \sqrt{\lambda_e \chi_{1,0.05}^2} = -\sqrt{0.00393\, \lambda_e},
\qquad
\mathrm{cv}_{EM21,0.05}^c = \sqrt{\lambda_e \chi_{1,0.05}^2} = \sqrt{0.00393\, \lambda_e},
\]
at significance level $0.05$ (corresponding to confidence level $0.9$).

Next, we study asymptotic behavior under the alternative.

\begin{theorem}
    \label{theorem:collapse:H1}
    Suppose Assumption~\ref{assumption:error} holds.
    \begin{itemize}
    \item[(i)] If $T_c > T_1$ ($\lambda_c > \lambda_1$), then
    \[
    \mathrm{LR}_{a,12}^c,\; \mathrm{EM}_{a,12}^c,\; \mathrm{EM}_{b,12}^c \to \infty,
    \quad \text{whereas} \quad
    \mathrm{LR}_{a,21}^c,\; \mathrm{EM}_{a,21}^c,\; \mathrm{EM}_{b,21}^c = O_p(1).
    \]
    \item[(ii)] If $T_c < T_1$ ($\lambda_c < \lambda_1$), then
    \[
    \mathrm{LR}_{a,12}^c,\; \mathrm{EM}_{a,12}^c,\; \mathrm{EM}_{b,12}^c \xrightarrow{p} 0,
    \qquad
    \mathrm{LR}_{a,21}^c,\; \mathrm{EM}_{a,21}^c,\; \mathrm{EM}_{b,21}^c \xrightarrow{p} 0.
    \]
    \end{itemize}
\end{theorem}

From Theorem~\ref{theorem:collapse:H1}(i), the $\mathrm{LR}_{a,12}^c$, $\mathrm{EM}_{a,12}^c$, and $\mathrm{EM}_{b,12}^c$ tests are consistent against $T_c > T_1$ (right-tail rejection), whereas $\mathrm{LR}_{a,21}^c$, $\mathrm{EM}_{a,21}^c$, and $\mathrm{EM}_{b,21}^c$ are not. In contrast, Theorem~\ref{theorem:collapse:H1}(ii) implies that the latter tests are consistent against $T_c < T_1$ (left-tail rejection), while the former also reject because their right-tail critical values are negative. Overall, each test is consistent at least for the direction it is designed to detect, mirroring the emergence-date case.

As in the emergence-date confidence sets, Theorem~\ref{theorem:collapse:H1} suggests that sets based on $\mathrm{LR}_{a,12}^c$, $\mathrm{EM}_{a,12}^c$, and $\mathrm{EM}_{b,12}^c$ will include only a small fraction of permissible dates earlier than the true $T_c$, whereas sets based on $\mathrm{LR}_{a,21}^c$, $\mathrm{EM}_{a,21}^c$, and $\mathrm{EM}_{b,21}^c$ will tend to include relatively few dates later than $T_c$.

\subsection{Confidence Set for the Recovery Date of a Bubble}

In this subsection, we focus on the recovery date of a bubble in the regime $[T_c+1, T]$, and consider a one-break model:
\begin{align}
    y_t &=
    \begin{cases}
    \phi_b y_{t-1} + \varepsilon_t, & T_c + 1 \le t \le T_r,\\
    y_{t-1} + \varepsilon_t, & T_r + 1 \le t \le T.
    \end{cases}
    \label{model:recovery}
\end{align}
Let $T_{CU} \coloneqq T - T_c$ be the sample size of these two regimes.

\subsubsection{Confidence Set for the Recovery Date}

The limiting distribution of $\hat{T}_r$ is derived in Theorem~3 of \citet{KurozumiS2023} for $\alpha > \beta$, but it depends on the localizing parameter $b$ in a complicated way that cannot be consistently estimated. Hence, it is difficult to construct a confidence interval based on the limiting distribution of the breakpoint estimator.

Instead, as in the cases of the emergence and collapse dates, we construct the confidence set by inverting a test for
\[
H_0:\; T_r = T_1 \qquad \text{vs.} \qquad H_1:\; T_r = T_2 \ne T_1.
\]
Assuming $\{\varepsilon_t\} \sim i.i.d.\ N(0,\sigma^2)$, the likelihood principle implies rejection of $H_0$ when
\begin{align}
    2\sum_{t=T_1+1}^{T_2} y_{t-1}\Delta y_t + \rho_b \sum_{t=T_1+1}^{T_2} y_{t-1}^2 &< c_1 
    \quad \text{for } T_1 < T_2,
    \label{LR12:org:r}\\
    2\sum_{t=T_2+1}^{T_1} y_{t-1}\Delta y_t + \rho_b \sum_{t=T_2+1}^{T_1} y_{t-1}^2 &> c_2 
    \quad \text{for } T_2 < T_1,
    \label{LR21:org:r}
\end{align}
where $c_1$ and $c_2$ are determined by the significance level. We propose the following statistics:
\begin{align}
    \mathrm{LR}_{a,12}^{r}
    \coloneqq &
    \frac{\displaystyle \min_{T_2 \in \Lambda_{12}^r} 
      \left(
        2\sum_{t=T_1+1}^{T_2} y_{t-1}\Delta y_t
        + \hat{\rho}_b \sum_{t=T_1+1}^{T_2} y_{t-1}^2
      \right)}
      {\displaystyle T \,(T_{LRa12}^r - T_1)\, \hat{\rho}_b \,\hat{\phi}_a^{\,2(T_c - T_e)} \hat{\phi}_b^{\,2(T_1 - T_c)} \hat{\sigma}^2},
    \label{LRa12:r} \\
    \mathrm{LR}_{a,21}^{r}
    \coloneqq &
    \frac{\displaystyle \max_{T_2 \in \Lambda_{21}^r} 
      \left(
        2\sum_{t=T_2+1}^{T_1} y_{t-1}\Delta y_t
        + \hat{\rho}_b \sum_{t=T_2+1}^{T_1} y_{t-1}^2
      \right)}
      {\displaystyle T \,\hat{\phi}_a^{\,2(T_c - T_e)} \hat{\phi}_b^{\,2(T_{LRa21}^r - T_c)} \hat{\sigma}^2/2},
    \label{LRa21:r}    
\end{align}
where $T_{LRa12}^r$ minimizes the numerator of \eqref{LRa12:r}, 
$\Lambda_{12}^r \coloneqq \{T_2:\; T_1 + \lfloor T_{CU}\epsilon \rfloor \le T_2 \le T\}$, 
$T_{LRa21}^r$ maximizes the numerator of \eqref{LRa21:r}, and 
$\Lambda_{21}^r \coloneqq \{T_2:\; T_c + \lfloor T_{CU}\epsilon \rfloor \le T_2 \le T_1 - \lfloor T_{CU}\epsilon \rfloor\}$.
The $\mathrm{LR}_{a}^r$ test rejects $H_0$ if
\[
\mathrm{LR}_{a,12}^r < \mathrm{cv}_{LR12,\delta/2}^r
\qquad \text{or} \qquad
\mathrm{LR}_{a,21}^r > \mathrm{cv}_{LR21,1-\delta/2}^r,
\]
where $\mathrm{cv}_{LR12,\delta/2}^r$ and $\mathrm{cv}_{LR21,1-\delta/2}^r$ are the $\delta/2$ and $1-\delta/2$ quantiles of the respective null limiting distributions. 

We also consider a finite-sample correction for $\mathrm{LR}_{a,21}^r$:
\[
    \mathrm{LR}_{b,21}^r
    \coloneqq
    \frac{\displaystyle \max_{T_2 \in \Lambda_{21}^{r}}
      \left(
        - y_{T_2}^2 - \sum_{t=T_2+1}^{T_1} (\Delta y_t)^2
        + \hat{\rho}_b \sum_{t=T_2+1}^{T_1} y_{t-1}^2
      \right)}
    {\displaystyle T \,\hat{\phi}_a^{\,2(T_c - T_e)} \hat{\phi}_b^{\,2(T_{LRb21}^r - T_c)} \hat{\sigma}^2/2},
\]
where $T_{LRb21}^r$ maximizes the numerator. We will use $\mathrm{LR}_{b,21}^r$ in conjunction with $\mathrm{EM}_{b,12}^r$ introduced below; we refer to this combination as the $\mathrm{LE}^r$ test.

Proceeding as before, we can also base EM-type tests on $t$-statistics. We propose
\begin{equation}
\mathrm{EM}_{a,12}^r \coloneqq \frac{1}{T_{CU}} \sum_{T_2 = T_1 + \lfloor T_{CU}\epsilon \rfloor}^{T} t_{T_1,T_2}
\quad \text{and} \quad
\mathrm{EM}_{a,21}^r \coloneqq 
\frac{\displaystyle \sum_{T_2 = T_c + \lfloor T_{CU}\epsilon \rfloor}^{T_1 - \lfloor T_{CU}\epsilon \rfloor} t_{T_2,T_1}}
     {\sqrt{T \,\hat{\phi}_a^{\,2(T_c - T_e)} \hat{\phi}_{b}^{\,2\lfloor T_{CU}\epsilon \rfloor}/(2\hat{\rho}_b)}},
\label{EMa:r}
\end{equation}
and reject when
\[
    \mathrm{EM}_{a,12}^r < \mathrm{cv}_{EMa12,\delta/2}^r
    \quad \text{or} \quad
    \mathrm{EM}_{a,21}^r > \mathrm{cv}_{EMa21,1-\delta/2}^r,
\]
where $\mathrm{cv}_{EMa12,\delta/2}^r$ and $\mathrm{cv}_{EMa21,1-\delta/2}^r$ are the respective critical values. We refer to this as the $\mathrm{EM}_a^r$ test. We also consider minimum- and maximum-type tests:
\begin{equation}
    \mathrm{EM}_{b,12}^r \coloneqq \min_{T_2 \in \Lambda_{12}^r} t_{T_1,T_2}
    \quad \text{and} \quad
    \mathrm{EM}_{b,21}^r \coloneqq 
    \frac{\displaystyle \max_{T_2 \in \Lambda_{21}^r} t_{T_2,T_1}}
         {\sqrt{T \,\hat{\rho}_b \hat{\phi}_a^{\,2(T_c - T_e)} \hat{\phi}_{b}^{\,2(T_{EMb21}^r - T_c)} / 2}},
    \label{EMb:r}
\end{equation}
where $T_{EMb21}^r$ maximizes the numerator. Reject when
\[
    \mathrm{EM}_{b,12}^r < \mathrm{cv}_{EMb12,\delta/2}^r
    \quad \text{or} \quad
    \mathrm{EM}_{b,21}^r > \mathrm{cv}_{EM21,\delta/2}^r.
\]
We refer to this as the $\mathrm{EM}_b^r$ test.

\subsubsection{Limiting Distributions of the Tests for the Recovery Date of a Bubble}

We first present the limiting distributions under the null. A complication is that the limiting distributions depend on whether $\alpha < \beta$ or $\alpha > \beta$.\footnote{The knife-edge case $\alpha=\beta$ is more involved (see \citet{KurozumiS2023}). As the behavior of the tests can be inferred from the analyses of $\alpha<\beta$ and $\alpha>\beta$,  we do not pursue it here.}

\begin{theorem}
    \label{theorem:recovery:H0}
    Suppose Assumption~\ref{assumption:error} holds. Under $H_0:\, T_r = T_1$,
    \begin{itemize}
    \item[(i)] If $\alpha < \beta$, then
    \begin{align}
        \mathrm{LR}_{a,12}^r &\Rightarrow W^2(\lambda_e),
        \label{limit:LR12:H0:recovery} \\
        \mathrm{EM}_{a,12}^r &\Rightarrow 
        \int_{\lambda_1^* + \epsilon}^{1}
        \frac{W(\lambda_2^*) - W(\lambda_1^*)}{\sqrt{\lambda_2^* - \lambda_1^*}}\, d\lambda_2^*,
        \label{limit:EMa12:H0:recovery} \\
        \mathrm{EM}_{b,12}^r &\Rightarrow 
        \inf_{\lambda_1^* + \epsilon < \lambda_2^* < 1}
        \frac{W(\lambda_2^*) - W(\lambda_1^*)}{\sqrt{\lambda_2^* - \lambda_1^*}},
        \label{limit:EMb12:H0:recovery}
    \end{align}
    where $\lambda_j^* \coloneqq (T_j - T_c)/T_{CU}$ for $j=1,2$, and
    \begin{align}
        \mathrm{LR}_{a,21}^r &\Rightarrow - W^2(\lambda_e),
        \label{limit:LRa21:H0:recovery} \\
        \mathrm{EM}_{a,21}^r,\; \mathrm{EM}_{b,21}^r &\Rightarrow - |W(\lambda_e)|.
        \label{limit:EM21:H0:recovery}
    \end{align}
    \item[(ii)] If $\alpha > \beta$, then
    \begin{align}
        \mathrm{LR}_{a,12}^r &\to \infty, \nonumber \\
        \mathrm{EM}_{a,12}^r &\Rightarrow 
        \int_{\lambda_1^* + \epsilon}^{1}
        ADF^r(\lambda_1^*, \lambda_2^*)\, d\lambda_2^*,
        \label{limit:EMa12s:H0:recovery} \\
        \mathrm{EM}_{b,12}^r &\Rightarrow 
        \inf_{\lambda_1^* + \epsilon < \lambda_2^* < 1}
        ADF^r(\lambda_1^*, \lambda_2^*),
        \label{limit:EMb12s:H0:recovery}
    \end{align}
    where
    \[
    ADF^r(\lambda_1^*, \lambda_2^*) \coloneqq
    \frac{\frac{1}{2}\left[(W(\lambda_2^*) - W(\lambda_1^*))^2 - (\lambda_2^* - \lambda_1^*)\right]}
         {\sqrt{\int_{\lambda_1^*}^{\lambda_2^*} (W(s) - W(\lambda_1^*))^2 ds}},
    \]
    and
    \[
        \mathrm{LR}_{a,21}^r,\;
        \mathrm{EM}_{a,21}^r,\;
        \mathrm{EM}_{b,21}^r \to -\infty.
    \]
    \end{itemize}
\end{theorem}

We must choose critical values in light of Theorem~\ref{theorem:recovery:H0}. For $\mathrm{LR}_{a,12}^r$, a nominal level (say, $0.05$) is asymptotically controlled by rejecting when
\[
\mathrm{LR}_{a,12}^r < \mathrm{cv}_{LR12,0.05}^r
= \lambda_e \chi_{1,0.05}^2
= 0.00393\,\lambda_e,
\]
if $\alpha < \beta$. Because the rejection region is the left tail, the test becomes conservative (the theoretical size tends to zero) when $\alpha > \beta$ if we use the above critical values. Note that the numerator of $\mathrm{LR}_{a,12}^r$ is $O_p(T^{2-\beta})$ when $\alpha > \beta$ (see the proof of Theorem~\ref{theorem:recovery:H0}). Thus, scaling the LR-type statistic by $T^{2-\beta}$ would control size when $\alpha > \beta$, but would drive the statistic to zero in probability when $\alpha < \beta$, yielding a liberal test (size tends to one). To ensure conservativeness uniformly, we adopt the critical values based on \eqref{limit:LR12:H0:recovery}.

By contrast, $\mathrm{EM}_{a,12}^r$ and $\mathrm{EM}_{b,12}^r$ converge in distribution in both cases. Simulations indicate that the critical values based on \eqref{limit:EMa12s:H0:recovery} and \eqref{limit:EMb12s:H0:recovery} are smaller than those based on \eqref{limit:EMa12:H0:recovery} and \eqref{limit:EMb12:H0:recovery}, respectively. Hence, we use the former as $\mathrm{cv}_{EMa12,\delta/2}^r$ and $\mathrm{cv}_{EMb12,\delta/2}^r$. Because the limits depend on $\lambda_1^*$, we fit the response-surface regression
\[
\mathrm{cv}_{EMa12,0.05}^r
= a_{0,\ell} + a_{-1,\ell}\frac{1}{\lambda_1^*} + a_{1,\ell}\lambda_1^* + a_{2,\ell}\lambda_1^{*2} + a_{3,\ell}\lambda_1^{*3},
\]
and, for $\mathrm{EM}_{b,12}^r$,
\begin{align*}
    \mathrm{cv}_{EMb12,0.05}^r
    &= a_{0,\ell} + a_{-1,\ell}\frac{1}{\lambda_1^*} + a_{1,\ell}\lambda_1^* + a_{2,\ell}\lambda_1^{*2} + a_{3,\ell}\lambda_1^{*3} \\
    &\quad + \mathbf{1}\{\lambda_1^* > 0.7\}
    \left[
      b_{0,\ell} + b_{-1,\ell}\frac{1}{\lambda_1^*} + b_{1,\ell}\lambda_1^* + b_{2,\ell}\lambda_1^{*2} + b_{3,\ell}\lambda_1^{*3}
    \right].
\end{align*}
The estimated coefficients are reported in Table~1.

Similarly, the critical values based on \eqref{limit:LRa21:H0:recovery} and \eqref{limit:EM21:H0:recovery} yield conservative tests in general, so we use
\[
\mathrm{cv}_{LR21,0.95}^r = - \lambda_e \chi_{1,0.05}^2 = -0.00393\,\lambda_e,
\qquad
\mathrm{cv}_{EM21,0.95}^r = - \sqrt{\lambda_e \chi_{1,0.05}^2} = -\sqrt{0.00393\,\lambda_e}.
\]

Next, we study asymptotic behavior under the alternative.

\begin{theorem}
    \label{theorem:recovery:H1}
    Suppose Assumption~\ref{assumption:error} holds.
    \begin{itemize}
    \item[(i)] If $\alpha < \beta$:
        \begin{itemize}
        \item[(i-a)] If $T_r > T_1$ ($\lambda_r > \lambda_1$),
        \[
        \mathrm{LR}_{a,12}^r \to 0, 
        \qquad
        \mathrm{EM}_{a,12}^r,\; \mathrm{EM}_{b,12}^r \to -\infty,
        \]
        whereas
        \[
        \mathrm{LR}_{a,21}^r,\; \mathrm{LR}_{b,21}^r,\; \mathrm{EM}_{a,21}^r,\; \mathrm{EM}_{b,21}^r = O_p(1).
        \]
        \item[(i-b)] If $T_r < T_1$ ($\lambda_r < \lambda_1$),
        \[
        \mathrm{LR}_{a,12}^r \to \infty,
        \qquad
        \mathrm{EM}_{a,12}^r,\; \mathrm{EM}_{b,12}^r = O_p(1),
        \]
        whereas
        \[
        \mathrm{LR}_{a,21}^r,\; \mathrm{LR}_{b,21}^r \to \infty,
        \qquad
        \mathrm{EM}_{a,21}^r = O_p(1),
        \qquad
        \mathrm{EM}_{b,21}^r \xrightarrow{p} 0.
        \]
        \end{itemize}
    \item[(ii)] If $\alpha > \beta$:
        \begin{itemize}
        \item[(ii-a)] If $T_r > T_1$ ($\lambda_r > \lambda_1$),
        \[
        \mathrm{LR}_{a,12}^r,\; \mathrm{EM}_{a,12}^r,\; \mathrm{EM}_{b,12}^r \to -\infty,
        \]
        whereas
        \[
        \mathrm{LR}_{a,21}^r,\; \mathrm{LR}_{b,21}^r \to \infty,
        \qquad
        \mathrm{EM}_{a,21}^r,\; \mathrm{EM}_{b,21}^r \to \infty \text{ or } -\infty.
        \]
        \item[(ii-b)] If $T_r < T_1$ ($\lambda_r < \lambda_1$),
        \[
        \mathrm{LR}_{a,12}^r\to \infty,
        \qquad
         \mathrm{EM}_{a,12}^r,\; \mathrm{EM}_{b,12}^r = O_p(1),
        \]
        whereas
        \[
        \mathrm{LR}_{a,21}^r,\; \mathrm{LR}_{b,21}^r \to \infty,
        \qquad
        \mathrm{EM}_{a,21}^r,\; \mathrm{EM}_{b,21}^r \to \infty \text{ or } -\infty.
        \]
        \end{itemize}
    \end{itemize}
\end{theorem}

Interpreting Theorem~\ref{theorem:recovery:H1} requires care. From (i-a) and (ii-a), $\mathrm{LR}_{a,12}^r$, $\mathrm{EM}_{a,12}^r$, and $\mathrm{EM}_{b,12}^r$ are consistent against the designated alternative $T_r = T_2 > T_1$. By contrast, while $\mathrm{LR}_{a,21}^r$ and $\mathrm{LR}_{b,21}^r$ are consistent against $T_r = T_2 < T_1$ regardless of whether $\alpha < \beta$ or $\alpha > \beta$, $\mathrm{EM}_{b,21}^r$ is consistent against this alternative only when $\alpha < \beta$, and $\mathrm{EM}_{a,21}^r$ is inconsistent in both cases. Such inconsistencies would theoretically inflate the confidence set, although, as we show in the next section, finite-sample behavior need not perfectly mirror these asymptotic patterns.

\begin{remark}
\label{remark:joint}
Constructing a joint confidence set for the three break dates is more challenging than constructing the confidence sets separately. One difficulty is that the test statistic for the locations of the three breaks becomes more complicated, because there are eight possible combinations of break locations under the null and alternative hypotheses. Let $T_1^e$, $T_2^e$, $T_1^c$, $T_2^c$, $T_1^r$, and $T_2^r$ denote the emergence, collapse, and recovery dates under the null and alternative hypotheses, respectively. The test must then account for the combinations $T_1^e \gtrless T_2^e$, $T_1^c \gtrless T_2^c$, and $T_1^r \gtrless T_2^r$, which complicates formal inference.

A simple ad hoc approach is to use the test statistics developed separately for the confidence sets for $T_e$, $T_c$, and $T_r$. For example, one may apply the LR-type tests for the three breaks and reject the joint null hypothesis $T_e=T_1^e$, $T_c=T_1^c$, and $T_r=T_1^r$ if at least one of the $\mbox{LR}^e$, $\mbox{LR}^c$, or $\mbox{LR}^r$ tests rejects. To control the overall size at $\delta$, the significance level for each test must be set to $\delta/3$ (or $\delta/6$ for each one-sided test). However, using smaller significance levels reduces power and therefore tends to produce wider confidence sets. Moreover, all permissible combinations of $T_1^e$, $T_1^c$, and $T_1^r$ must be considered, which makes the procedure computationally demanding. For these reasons, we focus on constructing the confidence sets separately and leave the joint confidence set for future work.
\end{remark}

\section{Monte Carlo Simulations}

In this section, we investigate the finite-sample properties of the confidence sets for the emergence, collapse, and recovery dates of a bubble proposed in the previous sections. The data-generating process is given by \eqref{model:emerging} with $\{\varepsilon_t\} \sim i.i.d.\ N(0, 6.79^2)$, $y_0 = 100$, and $T = 200$, which follows a setting similar to that considered in PSY. The AR(1) parameters are specified as local-to-unity: $\phi_a = 1 + a/T$ and $\phi_b = 1 - a/T$ with $a \in \{2,4,6\}$, referred to as the small, medium, and large bubble, respectively. We report the empirical coverage rate and the average length of the confidence set relative to the regime length used in the tests for the break-date location. Results are based on 2{,}000 replications with a nominal confidence level of 0.9, implemented by setting the significance level to 0.05 for each one-sided location test. Note that we refer to the latter as ``length,'' even though the confidence set can be a (possibly) discontinuous subset of sample points. We also note that in the simulations, we always include the point estimate of the break date in the confidence set even though it might be possible for it to be excluded by the tests proposed in the previous section.

We consider the following break-date specifications: (Case 1) $\lambda_e = 0.3$, $\lambda_c = 0.5$, and $\lambda_r = 0.7$; (Case 2) $\lambda_e = 0.4$, $\lambda_c = 0.6$, and $\lambda_r = 0.8$; (Case 3) $\lambda_e = 0.5$, $\lambda_c = 0.7$, and $\lambda_r = 0.9$; and (Case 4) $\lambda_e = 0.4$, $\lambda_c = 0.6$, and $\lambda_r = 0.7$. We report the detailed results for Case 1; the remaining cases are summarized in the online appendix.

To construct the confidence set, we must prespecify the two consecutive regimes within which the true break date may be included. We conduct simulations under two settings: when the endpoints of the two regimes are correctly specified and when they are based on point estimates. In the former case, we construct the confidence sets for $T_e$, $T_c$, and $T_r$ over the intervals $[1,\;T_c]$, $[T_e+1,\;T_r]$, and $[T_c+1,\;T]$, respectively. In the latter case, to reduce the possibility that two break dates are included in the same interval due to potential estimation errors in $\hat{T}_{e}$, $\hat{T}_{c}$, and $\hat{T}_r$, we construct the confidence set over the trimmed intervals $[1,\;\hat{T}_c^{-}]$, $[\hat{T}_e^{+}+1,\;\hat{T}_r^{-}]$, and $[\hat{T}_c^{+}+1,\;T]$, respectively, where $\hat{T}_c^-\coloneqq \hat{T}_c-\lfloor 0.05\hat{T}_c\rfloor$, $\hat{T}_e^+\coloneqq \hat{T}_e+\lfloor 0.05(\hat{T}_r-\hat{T}_e)\rfloor$, $\hat{T}_r^-\coloneqq \hat{T}_r-\lfloor 0.05(\hat{T}_r-\hat{T}_e)\rfloor$, and $\hat{T}_c^{+}\coloneqq \hat{T}_c+\lfloor 0.05(T-\hat{T}_c)\rfloor$.

Table~\ref{table:Te:case1:true} summarizes the results for the confidence set of $T_e$ when the sample period is correctly specified as $t = 1, \ldots, T_c$. Results using the estimated end of the sample are presented later. The rows labeled ``coverage'' report the empirical coverage rates, which should equal 0.9 in theory, whereas ``coverage12'' (``coverage21'') corresponds to the coverage rates of the one-sided tests $\mathrm{LR}_{\cdot,12}^e$ and $\mathrm{EM}_{\cdot,12}^e$ ($\mathrm{LR}_{\cdot,21}^e$ and $\mathrm{EM}_{\cdot,21}^e$), each of which should be 0.95 theoretically. For example, when $a = 2$, the coverage rates of $\mathrm{LR}_a^e$, $\mathrm{LR}_{a,12}^e$, and $\mathrm{LR}_{a,21}^e$ are 0.19, 0.39, and 0.56, respectively. We see that $\mathrm{LR}_{\cdot,12}^e$ and $\mathrm{EM}_{\cdot,12}^e$ under-cover the true emergence date for small $a$, with coverage improving as $a$ increases. Among the one-sided tests, $\mathrm{LR}_{b,12}^e$ (appearing at the ``coverage12'' rows in the ``$\mathrm{LE}^e$'' column) and $\mathrm{EM}_{a,21}^e$ deliver better coverage, although the former is slightly conservative; the latter's coverage is very close to the nominal 0.95. Motivated by these properties, we combine $\mathrm{LR}_{b,12}^e$ and $\mathrm{EM}_{a,21}^e$ as $\mathrm{LE}^e$, which achieves coverage close to the nominal 90\%.

Average lengths of the confidence sets appear in the lower panel of Table~\ref{table:Te:case1:true}. The rows labeled ``length'' report the fraction of observations included in the confidence set relative to $T_c$, the sample size used for the tests. In general, a more liberal test yields a shorter relative length, which we observe in Table~\ref{table:Te:case1:true}. For the small bubble, the confidence set based on $\mathrm{LE}^e$ is the largest among all methods---about 61\% of the sample from $1$ to $T_c$---yet it is the only method that controls coverage in this case. As the explosive speed increases, the confidence set shortens; for the medium bubble, $\mathrm{LE}^e$ is comparable to $\mathrm{EM}_a^e$ (the second-best in terms of coverage), and for the large bubble, $\mathrm{LE}^e$ yields shorter sets than $\mathrm{EM}_a^e$.

To further examine the finite-sample behavior of each one-sided location test, we decompose the confidence set into two parts: the ``left-hand side'' consists of dates in the set less than or equal to the true emergence date, and the ``right-hand side'' consists of dates greater than or equal to $T_e$. The rows ``length12left'' and ``length12right'' report average lengths of the left- and right-hand portions for $\mathrm{LR}_{\cdot,12}^e$ and $\mathrm{EM}_{\cdot,12}^e$, while ``length21left'' and ``length21right'' report the corresponding quantities for $\mathrm{LR}_{\cdot,21}^e$ and $\mathrm{EM}_{\cdot,21}^e$. Since each one-sided test is consistent in its designated direction, we expect the left-hand (right-hand) portion from $\mathrm{LR}_{\cdot,12}^e$ and $\mathrm{EM}_{\cdot,12}^e$ ($\mathrm{LR}_{\cdot,21}^e$ and $\mathrm{EM}_{\cdot,21}^e$) to shrink as explosiveness increases, provided coverage is controlled. This pattern is borne out in finite samples. For instance, the left-hand length for $\mathrm{LE}_{12}^e$ is 0.48, 0.39, and 0.29 for $a = 2, 4, 6$, respectively, while the right-hand length for $\mathrm{LE}_{21}^e$ is 0.29, 0.20, and 0.14, with coverage well controlled in each case.

Table~\ref{table:Te:case1:est} reports coverage and lengths when the sets are constructed using the estimated subsample $t = 1, \ldots,\hat{T}_c^-$. Here, the right endpoint can fall before $T_e$, in which case reporting the length is not meaningful. Thus, average lengths are computed only for replications with $T_e\leq \hat{T}_c^-$, whereas coverage rates are computed over all replications. In this design, all methods yield liberal coverage when $a = 2$, partly because the estimated collapsing date precedes the true emerging date in 16\% of the 2{,}000 replications. This occurs in only 6\% and 2\% of replications for the medium and large bubbles, respectively, and the overall patterns then approach those in Table~\ref{table:Te:case1:true}.

Table~\ref{table:Tc:case1:true} summarizes the results for the confidence set of $T_c$ when the sample period is correctly specified as $t = T_e + 1, \ldots, T_r$. For the small bubble, all tests under-cover the true collapsing date; $\mathrm{EM}_a^c$ performs best among the four but still attains only 0.76, below the nominal 0.9. As $a$ increases, performance improves. For the medium bubble, $\mathrm{LR}_a^c$ and $\mathrm{EM}_b^c$ are liberal but closer to nominal (0.82), whereas $\mathrm{EM}_a^c$and $\mathrm{LE}^c$ (the combination of $\mbox{LR}_{a,21}^c$ and $\mbox{EM}_{a,12}^c$) are conservative (0.95 and 0.91). For the large bubble, all tests are conservative.

Regarding the length of the confidence set, $\mathrm{EM}_a^c$ delivers the longest sets, which is partly the price for improved coverage. Consistent with Section~3.6, the sets based on $\mathrm{LR}_{a,12}^c$, $\mathrm{EM}_{a,12}^c$, and $\mathrm{EM}_{b,12}^c$ tend to include only a small portion of dates earlier than the true $T_c$, whereas $\mathrm{LR}_{a,21}^c$, $\mathrm{EM}_{a,21}^c$, and $\mathrm{EM}_{b,21}^c$ yield sets with relatively few dates later than $T_c$.

Table~\ref{table:Tc:case1:est} reports coverage and lengths when sets are constructed using the estimated break-interval $[\hat{T}_e^{+}+1, \hat{T}_r^{-}]$. As in the emergence-date case, average lengths are computed only for replications with $T_c \in [\hat{T}_e^{+}+1, \hat{T}_r^{-}]$. All methods show smaller coverage rates here, partly because $T_c \notin [\hat{T}_e^{+}, \hat{T}_r^{-}]$ in 27\% of replications. For the medium bubble, the coverages of $\mathrm{EM}_a^c$ and $\mathrm{LE}^c$ are close to nominal (0.85 and 0.84), and for the large bubble they are slightly conservative (0.94). The lengths are similar to those under the true-ends design.

Tables~\ref{table:Tr:case1:true} and \ref{table:Tr:case1:est} present the results for $T_r$ when the sample period is correctly specified as $[T_c+1, T]$ and when the lower end is estimated as $[\hat{T}_c^{+}+1, T]$, respectively. We report results for $\mathrm{LR}_a^r$, $\mathrm{EM}_a^r$, $\mathrm{EM}_b^r$, and $\mathrm{LE}^r$ (the combination of $\mathrm{EM}_{b,12}^r$ and $\mathrm{LR}_{b,21}^r$). With true break ends, $\mathrm{LE}^r$ attains the best (slightly conservative) coverage of about 0.93--0.95. Its set length is the largest for the small bubble but comparable to $\mathrm{EM}_a^r$ for the medium bubble (whose coverage is 0.89, close to nominal), and shorter than $\mathrm{EM}_a^r$ for the large bubble. Using estimated ends, coverage decreases and becomes too liberal for the small bubble---again partly because $T_r \notin [\hat{T}_c^{+}+1, T]$ in 19\% of replications---but is closer to nominal for the medium and large bubbles.

\begin{figure}[h]
\begin{minipage}{.3\linewidth}
\centering
\includegraphics[width=0.95\linewidth]{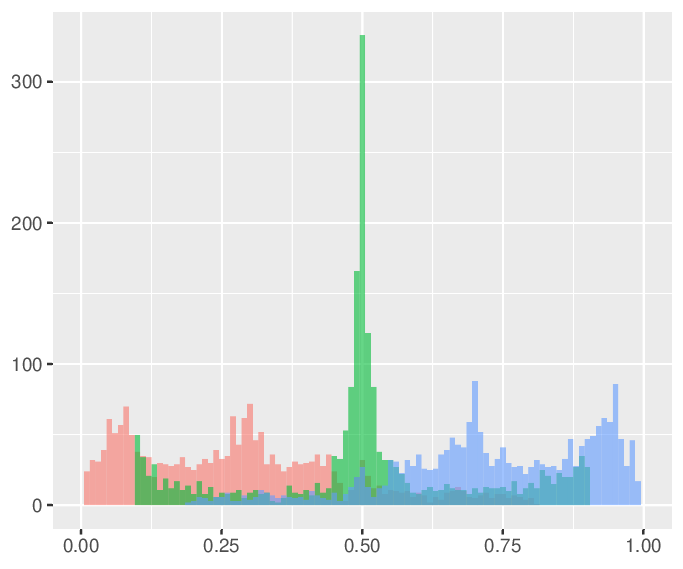}
(a) $a=2$
\end{minipage}
\begin{minipage}{.3\linewidth}
\centering
\includegraphics[width=0.95\linewidth]{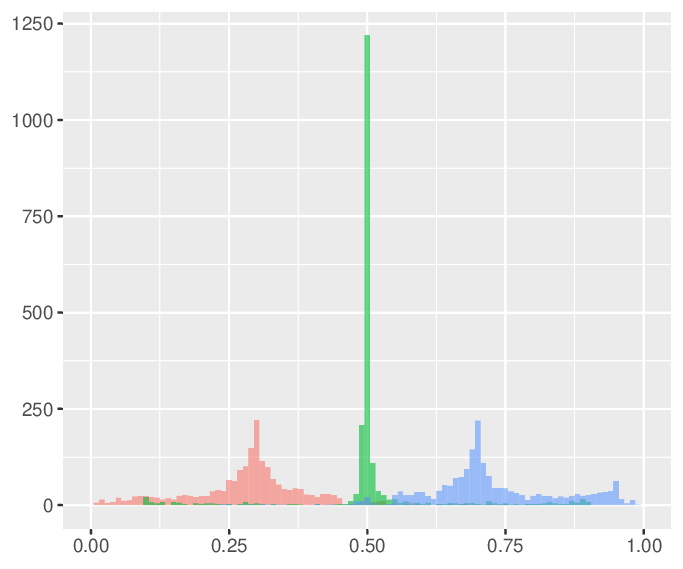}
(b) $a=4$
\end{minipage}
\begin{minipage}{.3\linewidth}
\centering
\includegraphics[width=0.95\linewidth]{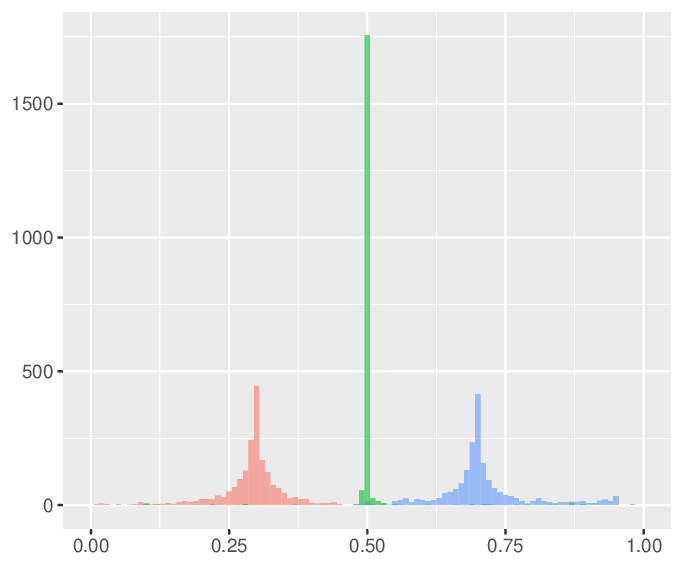}
(c) $a=6$
\end{minipage}
\caption{Finite Sample Distributions of the Break Points Estimates (Case 1)}
\medskip

\noindent
Note: The pink, green, and blue bins correspond to $\hat{T}_e$, $\hat{T}_c$, and $\hat{T}_r$, respectively. 
\label{figure:finite}
\end{figure}

\begin{figure}[h]
\begin{minipage}{.5\linewidth}
\centering
\includegraphics[width=0.95\linewidth]{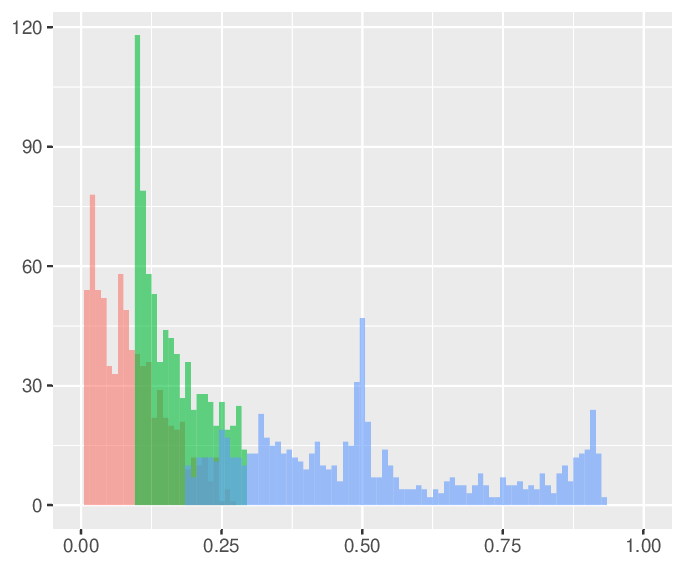}
(a) $a=2$, $\hat{T}_c < T_e$
\end{minipage}
\begin{minipage}{.5\linewidth}
\centering
\includegraphics[width=0.95\linewidth]{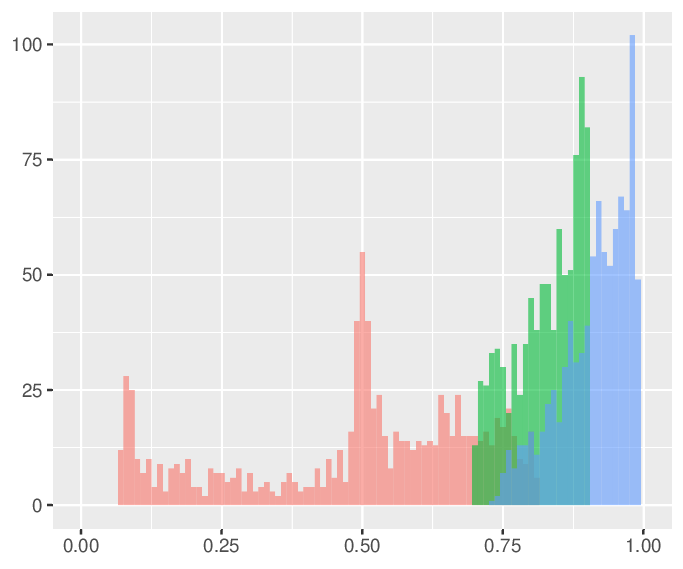}
(a-i) $a=2$, $T_r < \hat{T}_c$
\end{minipage}
\caption{Finite Sample Distributions of the Break Points Estimates When $\hat{T}_c < T_e$ or $T_r < \hat{T}_{c}$}
\medskip

\noindent
Note: The pink, green, and blue bins correspond to $\hat{T}_e$, $\hat{T}_c$, and $\hat{T}_r$, respectively. 
\label{figure:finite:limited}
\end{figure}

As the misidentification of bubble break dates appears to be nontrivial, we further investigate the finite-sample distributions of the break-date estimators. Figure \ref{figure:finite} presents histograms of the three estimators for $a=2$, 4, and 6, with the pink, green, and blue bins corresponding to $\hat{T}_e$, $\hat{T}_c$, and $\hat{T}_r$, respectively. The distributions are well separated when $a=6$, but overlap and are not unimodal when $a=2$. To examine the case $a=2$ more closely, we rerun the simulation with 5,000 replications instead of 2,000 and focus on the subsamples satisfying $\hat{T}_c < T_e$ and $T_r < \hat{T}_c$. Figure \ref{figure:finite:limited} reports the corresponding histograms. Figure \ref{figure:finite:limited}(a) shows that, conditional on $\hat{T}_c < T_e$, $\hat{T}_e$ and $\hat{T}_c$ tend to lie near the left endpoints of the permissible break-date ranges, whereas $\hat{T}_r$ peaks at the collapse date (0.5). By contrast, when $T_r < \hat{T}_c$, $\hat{T}_c$ and $\hat{T}_r$ tend to lie near the right endpoints of the permissible break-date ranges, whereas $\hat{T}_e$ peaks at the collapse date. These findings suggest that, in practical applications, if either $\hat{T}_e$ or $\hat{T}_r$ is estimated around the collapse date, which may be confirmed by visual inspection, while the other two break dates are estimated near the endpoints of the permissible ranges, the point estimates may be unreliable and the possible break dates should be examined more carefully.

To summarize the finite-sample results, $\mathrm{LE}^e$ and $\mathrm{LE}^r$ deliver well-controlled coverage with reasonable set lengths for medium and large bubbles. For the collapse date, no single test performs best in terms of both coverage rate and power. The coverage rate of $\mathrm{LE}^c$ is relatively close to 0.9 for the medium and large bubbles, while the length of its confidence set is the second shortest.

\section{Empirical Application}

\begin{figure}[h]
\centering
\includegraphics[width=0.8\linewidth]{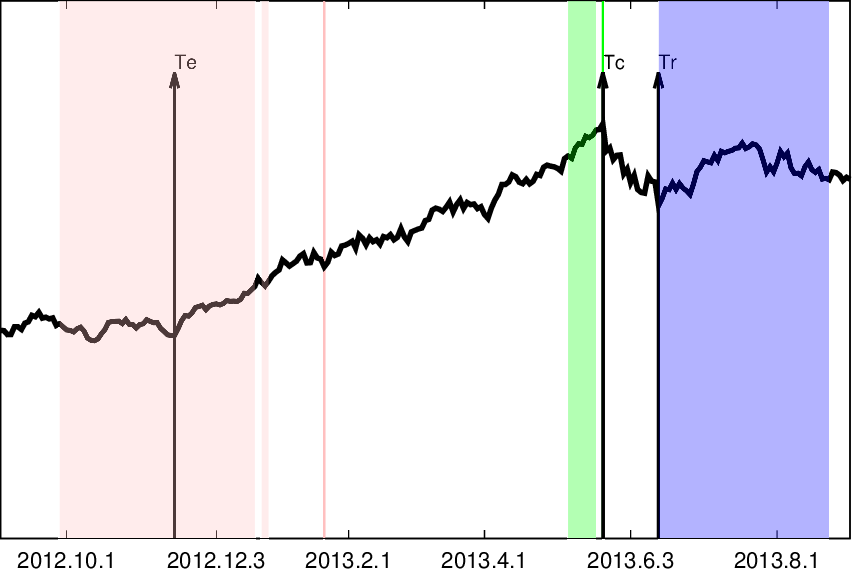}
\caption{Logarithm of Japanese stock price}
\medskip

\noindent
Note: The vertical arrows show the point estimates of $T_e$, $T_c$, and $T_r$, respectively, while the pink, green, and bue areas correspond to the confidence set for $T_e$, $T_c$, and $T_r$, respectively.
\label{figure:stock}
\end{figure}

In this section, we illustrate how to construct confidence sets for the emergence, collapse, and recovery dates of explosive behavior by applying the proposed methods to stock prices. Figure~\ref{figure:stock} plots the logarithm of daily Japanese stock prices (Nikkei~225) from September 2012 to August 2013.\footnote{The data that support the findings of this study are available from the corresponding author, Eiji Kurozumi, upon reasonable request.}  A visual inspection suggests mild explosiveness followed by a collapse over this period. The exuberance in stock prices may be associated with changes in the Bank of Japan (BOJ) monetary policy: the BOJ announced a 2\% inflation target in January 2013 and implemented quantitative and qualitative monetary easing in April 2013.

We first test for bubbles and reject the null of no bubble using the SADF test of \citet{PhillipsWY2011a} and the UR test of \citet{KurozumiNishi2024} at the 5\% significance level. Given evidence of exuberance, we estimate the emergence, collapse, and recovery dates by the sample-splitting approach of \citet{KurozumiS2023}. The resulting point estimates are November~14, 2012; May~22, 2013; and June~13, 2013, respectively, shown as vertical arrows---$T_e$, $T_c$, and $T_r$---in Figure~ref{figure:stock}. Although the estimated emergence predates the BOJ policy changes, our focus is whether the confidence sets include the dates of these policy shifts. Note that our method provides the confidence sets, which is more informative than the point estimates obtained by the existing method.

To this end, we construct a 90\% confidence set for $T_e$ based on the $\mathrm{LE}^e$ test introduced above. This set is depicted as the pink band in Figure~\ref{figure:stock}, which includes 61 sample points.\footnote{We can also construct a confidence interval for $T_e$ based on the limiting distribution of the break-date estimator, as considered in Section 3.1.1. However, this confidence interval ranges from September 28, 2012, to April 16, 2013, covering about 81\% of the permissible break dates. In contrast, the confidence interval based on $\mbox{LE}^e$ has a length of about 37\%, so our new method reduces the size of the confidence set by 44 percentage points.} Although early April are excluded from the confidence set, January 23, which is the next day of the announcement of the introduction of the inflation target, is included in the confidence set, suggesting that prices may begin rising just after the policy change.  We also note that the confidence set for the collapse date based on $\mathrm{LE}^c$ (green band) is narrow compared to the other confidence sets, with only 10 sample points. Finally, the blue band shows the confidence set for $T_r$ based on $\mathrm{LE}^r$. We conclude that it is unlikely that a reversion to normal market behavior occurred earlier than the point estimate of $T_r$, whereas the market recovery date might be much later than $\hat{T}_r$ as the confidence set includes 49 sample points later than $\hat{T}_r$.

\section{Concluding Remarks}

We propose constructing confidence sets for the emergence, collapse, and recovery dates of a bubble, separately, by inverting tests for the location of the break date: one class based on the likelihood ratio (LR) principle and two Elliott--M\"uller-type tests \citep{ElliottM2007}. We also introduce finite-sample modifications aimed at better controlling the empirical confidence level. In general, coverage probabilities are well controlled for large bubbles, whereas under-coverage tends to arise under weak explosiveness. Monte Carlo evidence further shows that, the faster the explosive speed, the shorter the resulting confidence set.

Our analysis is based on model \eqref{model:0}, but there are several alternative ways to model the collapsing regime. Such alternatives would require different constructions for the confidence sets of the collapsing and recovery dates; this is the subject of our ongoing research.

\bibliographystyle{apalike} 
\bibliography{ref_bubble}

@article{ElliottM2007,
  title   = {Confidence Sets for the Date of a Single Break in Linear Time Series Regressions},
  author  = {Elliott, G. and M\"uller, U. K.},
  journal = {Journal of Econometrics},
  volume  = {141},
  pages   = {1196--1218},
  year    = {2007}
}

@article{EoM2015,
  title   = {Likelihood-Ratio-Based Confidence Sets for the Timing of Structural Breaks},
  author  = {Eo, Y. and Morley, J.},
  journal = {Quantitative Economics},
  volume  = {6},
  pages   = {463--497},
  year    = {2015}
}

@article{Hansen2000b,
  title   = {Sample Splitting and Threshold Estimation},
  author  = {Hansen, B. E.},
  journal = {Econometrica},
  volume  = {68},
  pages   = {575--603},
  year    = {2000}
}

@article{HarveyLS2017,
  title   = {Improving the Accuracy of Asset Price Bubble Start and End Date Estimators},
  author  = {Harvey, D. I. and Leybourne, S. J. and Sollis, R.},
  journal = {Journal of Empirical Finance},
  volume  = {40},
  pages   = {121-138},
  year    = {2017}
}

@article{HarveyLST2016,
    author={Harvey, D. I. and Leybourne, S. J. and Sollis, R. and Taylor, A. M. R.},
    year={2016},
    title={Tests for Explosive Financial Bubbles in the Presence of Non-stationary Volatility},
    journal={Journal of Empirical Finance},
    volume={38},
    pages={548-574}
}

@article{HarveyLTZ2024,
  title   = {A New Heteroskedasticity‐Robust Test for Explosive Bubbles},
  author  = {Harvey, D. I. and Leybourne, S. J. and  Taylor, A. M. R. and Zu, Y.},
  journal = {Journal of Time Series Analysis},
  volume  = {46},
  pages   = {846-866},
  year    = {2025},
  note={\url{https://doi.org/10.1111/jtsa.12784}}
}

@article{HarveyLW2020,
    author={Harvey, D. I. and Leybourne, S. J. and Whitehouse, E. J.},
    year={2020},
    title={Date-stamping Multiple Bubble Regimes},
    journal={Journal of Empirical Finance},
    volume={58},
    pages={226-246}
}

@article{HarveyLZ2019,
  title   = {Testing Explosive Bubbles with Time-varying Volatility},
  author  = {Harvey, D. I. and Leybourne, S. J. and Zu, Y.},
  journal = {Econometric Reviews},
  volume  = {38},
  pages   = {1131-1151},
  year    = {2019},
  doi     = {10.1080/07474938.2018.1351669}
}

@article{HarveyLZ2020,
    author={Harvey, D. I. and Leybourne, S. J. and Zu, Y.},
    year={2020},
    title={Sign-based Unit Root Tests for Explosive Financial Bubbles in the Presence of Deterministically Time-varying Volatility},
    journal={Econometric Theory},
    volume={36},
    pages={122-169}
}

@article{KejriwalNP2025,
  title   = {An Improved Procedure for Retrospectively Dating the Emergence and Collapse of Bubbles},
  author  = {Kejriwal, M. and Nguyen, L. and Perron, P.},
  journal = {Journal of Time Series Analysis},
  volume  = {46},
  pages   = {867-883},
  year    = {2025},
  note={\url{https://doi.org/10.1111/jtsa.12810}}
}

@article{KH09,
  title   = {Asymptotic Properties of the Efficient Estimators for Cointegrating Regression Models with Serially Dependent Errors},
  author  = {Kurozumi, E. and Hayakawa, K.},
  journal = {Journal of Econometrics},
  volume  = {149},
  pages   = {118-135},
  year    = {2009},
}

@article{KurozumiNishi2024,
  title   = {Testing for a Bubble with a Stochastically Varying Explosive Coefficient},
  author  = {Kurozumi, E. and Nishi, M.},
  journal = {Journal of Time Series Analysis},
  volume  = {46},
  pages   = {945-965},
  year    = {2025},
  note={\url{https://doi.org/10.1111/jtsa.12768}}
}

@article{KurozumiSkrobotov2018,
    title={Confidence Sets for the Break Date in Cointegrating Regressions},
    author={Kurozumi, E. and Skrobotov, A. },
    journal={Oxford Bulletin of Economics and Statistics},
    volume ={80},
    pages  ={514-535},
    year={2018}
}

@article{KurozumiS2023,
    title={On the Asymptotic Behavior of Bubble Date Estimators},
    author={Kurozumi, E. and Skrobotov, A. },
    journal={Journal of Time Series Analysis},
    volume ={44},
    pages  ={359-373},
    year={2023},
    note={\url{https://doi.org/10.1111/jtsa.12672}}
}

@article{KurozumiST2023,
    author={Kurozumi, E. and Skrobotov, A. and Tsarev, A. },
    title={Time-Transformed Test for Bubbles under Non-stationary Volatility},
    journal={Journal of Financial Econometrics},
    volume = {21},
    pages = {1282-1307},
    year  = {2023},
    note={\url{https://doi.org/10.1093/jjfinec/nbac004}}
}

@article{KurozumiYamamoto2015,
  title   = {Confidence Sets for the Break Date Based on Optimal Tests},
  author  = {Kurozumi, E. and Yamamoto, Y.},
  journal = {Econometrics Journal},
  volume  = {18},
  pages   = {412-435},
  year    = {2015}
}

@article{PangCZL2018,
  title = {Structural Change in Nonstationary {AR} (1) Models},
  author = {Pang, T. and Chong, T. T-L. and Zhang, E. and Liang, Y.},
  journal	= {Econometric Theory},
  volume	= {34},
  pages = {985-1017},
  year = {2018}
}

@article{PangDC2021,
  title={Estimating multiple breaks in nonstationary autoregressive models},
  author={Pang, T. and Du, L. and Chong, T.T-L.},
  journal={Journal of Econometrics},
  volume={221},
  pages={277-311},
  year={2021}
}

@article{PhillipsS2018,
    author={Phillips, P. C. B. and Shi, S. },
    title={Financial Bubble Implosion and Reverse Regression},
    journal={Econometric Theory},
    volume={34},
    pages={705-753},
    year={2018}
}

@article{PhillipsSY2015a,
  title   = {Testing for Multiple Bubbles: Historical Episodes of Exuberance and Collapse in the {S\&P 500}},
  author  = {Phillips, P. C. B. and Shi, S. and Yu, J.},
  journal = {International Economic Review},
  volume  = {56},
  pages   = {1043-1077},
  year    = {2015}
}

@article{PhillipsSY2015b,
  title   = {Testing for Multiple Bubbles: Limit Theory of Real-Time Detectors},
  author  = {Phillips, P. C. B. and Shi, S. and Yu, J.},
  journal = {International Economic Review},
  volume  = {56},
  pages   = {1079-1133},
  year    = {2015}
}

@article{PhillipsWY2011a,
  title   = {Explosive Behavior in the 1990s {NASDAQ}: When Did Exuberance Escalate Asset Values?},
  author  = {Phillips, P. C. B. and Wu, Y. and Yu, J.},
  journal = {International Economic Review},
  volume  = {52},
  pages   = {201-226},
  year    = {2011}
}

@article{Skrobotov2023,
    author={Skrobotov, A. },
    title={Testing for Explosive Bubbles: A Review},
    journal={Dependence Modeling},
    year={2023},
    volume={11},
    note={\url{https://doi.org/10.1515/demo-2022-0152}}
}

@book{TANAKA96,
  title   = {Time Series Analysis: Nonstationary and Noninvertible Distribution Theory},
  author  = {Tanaka, K.},
  publisher = {Wiley},
  address   = {New York},
  year      = {1996},
}

@article{WuShiWu2024,
  title   = {Quantile Analysis for Financial Bubble Detection and Surveillance},
  author  = {Wu, R. and Shi, S. and Wu, J.},
  journal = {Journal of Time Series Analysis},
  volume  = {46},
  pages   = {908-931},
  year    = {2025},
  note={\url{https://doi.org/10.1111/jtsa.12791}}
}

@article{Yamamoto2018,
  title   = {A Modified Confidence Set for the Structural Break Date in Linear Regression Models},
  author  = {Yamamoto, Y.},
  journal = {Econometric Reviews},
  volume  = {37},
  pages   = {974-999},
  year    = {2018}
}

@article{Yamazaki2021,
  title   = {Improved Confidence Sets for the Date of a Structural Break},
  author  = {Yamazaki, D.},
  journal = {Econometric Reviews},
  volume  = {40},
  pages   = {257–289},
  year    = {2021}
}

\newpage

\newpage
\textbf{\Large{Appendix}}

Throughout the proofs, the notation $a_T \sim_a b_T$ indicates that $a_T / b_T \CP 1$ as $T \to \infty$.  
As shown in \citet{PangDC2021} and \citet{KurozumiS2023}, $\hat{\lambda}_e$ and $\hat{\lambda}_r$ are at least consistent for $\lambda_e$ and $\lambda_r$, respectively, while $\hat{T}_c$ is consistent for $T_c$. It then follows that
\[
\frac{\hat{\rho}_a}{\rho_a} \CP 1, \qquad
\frac{\hat{\phi}_a}{\phi_a} \CP 1, \qquad
\frac{\hat{\phi}_b}{\phi_b} \CP 1, \qquad
\text{and} \qquad
\hat{\sigma}^2 \CP \sigma^2.
\]
Therefore, in the following proofs, we replace $\hat{\rho}_a$, $\hat{\phi}_a$, $\hat{\phi}_b$, and $\hat{\sigma}^2$ with $\rho_a$, $\phi_a$, $\phi_b$, and $\sigma^2$, respectively, without loss of generality.

\noindent
\textbf{Proof of Theorem \ref{theorem:emergence:H0}}: We first note that, for $0\leq \lambda \leq \lambda_e$,
\begin{equation}
    \frac{1}{\sqrt{T}}y_{\lfloor T\lambda\rfloor}=\frac{1}{\sqrt{T}}\sum_{t=1}^{\lfloor T\lambda \rfloor} \varepsilon_t +o_p(1)\Rightarrow \sigma W(\lambda)
    \label{app:emergence:fCLT}
\end{equation}
by Assumption \ref{assumption:error}, whereas, as given in Lemma A1 in \citet{PhillipsSY2015b} and Lemma 2 in \citet{KurozumiS2023}, for $T_e+1\leq t \leq T_c$,
\begin{equation}
    y_t=\phi_a^{t-T_e}y_{T_e}(1+o_p(1)),
    \label{app:emergence:y}
\end{equation}
where the $o_p(1)$ term is uniform over $T_e< t \leq T_c$. In addition, we have
\begin{equation}
    \sum_{s=1}^{t}y_{s-1}\varepsilon_s=\left\{
    \begin{array}{lcl}
        O_p(T) & : & \mbox{for $t \leq T_e$} \\
        O_p(T^{\alpha+1}\phi_a^{t-T_e}) & : & \mbox{for $t>T_e$}
    \end{array},
    \right.
    \label{app:emergence:sum_ye}
\end{equation}
where the second result is obtained in Lemma A5 in \citet{PhillipsSY2015b} and Lemma 2 in \citet{KurozumiS2023}.

We first consider $\mbox{LR}_{\cdot 12}^e$ and $\mbox{EM}_{\cdot 12}^e$ for $T_1 < T_2$. Under the null hypothesis of $T_e=T_1$, we have, using \eqref{app:emergence:y},
\begin{align}
    \sum_{t=T_1+1}^{T_2}y_{t-1}^2
    & \sim_a
    y_{T_e}^2\sum_{t=T_1+1}^{T_2}\phi_a^{2(t-1-T_1)} \nonumber \\
    & \sim_a
    \frac{\sigma^2}{2}\frac{T\phi_a^{2(T_2-T_1)}}{\rho_a}\left(\frac{y_{T_1}^2}{T\sigma^2}\right),
    \label{app:emergence:sum_y2:T1T2}
\end{align}
and from \eqref{app:emergence:fCLT}--\eqref{app:emergence:sum_y2:T1T2},
\begin{align}
    \sum_{t=T_1+1}^{T_2}y_{t-1}\Delta y_t
    & =
    \frac{1}{2}\left\{ y_{T_2}^2-y_{T_1}^2-\sum_{t=T_1+1}^{T_2}(\Delta y_t)^2\right\} \nonumber \\
    & = 
    \frac{1}{2}\left\{ \phi_a^{2(T_2-T_1)}y_{T_e}^2(1+o_p(1))-y_{T_1}^2-\sum_{t=T_1+1}^{T_2}\left(\rho_ay_{t-1}+\varepsilon_t\right)^2\right\} \nonumber \\
    & \sim_a
    \frac{1}{2}\left( \phi_a^{2(T_2-T_1)}y_{T_1}^2-y_{T_1}^2-\rho_a^2\frac{\phi_a^{2(T_2-T_1)}}{2\rho_a}y_{T_1}^2\right) 
    \label{app:emergence:sum_ydy:T1T2pre} \\
    & \sim_a
    \frac{\sigma^2}{2}T\phi_a^{2(T_2-T_1)}\left(\frac{y_{T_1}^2}{T\sigma^2}\right),
    \label{app:emergence:sum_ydy:T1T2}
\end{align}
where the dominant term comes from $y_{T_2}^2$. Combining \eqref{app:emergence:sum_y2:T1T2} and \eqref{app:emergence:sum_ydy:T1T2}, we observe that
\[
2\sum_{t=T_1+1}^{T_2}y_{t-1}\Delta y_t-\rho_a\sum_{t=T_1+1}^{T_2}y_{t-1}^2
\sim_a
\frac{\sigma^2}{2}T\phi_a^{2(T_2-T_1)}\left(\frac{y_{T_1}^2}{T\sigma^2}\right).
\]
As the negligible terms absorbed in the symbol $\sim_a$ are uniform over $T_2\in \Lambda_{12}^{e}$, we have
\[
\mbox{LR}_{a,12}^e\Rightarrow W^2(\lambda_1).
\]

As $y_{T_2}^2$ is the dominant term in \eqref{app:emergence:sum_ydy:T1T2pre}, we can easily observe that $\mbox{LR}_{b,12}^e$ has the same weak limit as $\mbox{LR}_{a,12}^e$.

For $\mbox{EM}_{\cdot,12}^e$, we note from \eqref{app:emergence:sum_y2:T1T2} and \eqref{app:emergence:sum_ydy:T1T2} that
\begin{equation}
    t_{T_1,T_2}\sim_a \sqrt{\frac{T\rho_a}{2}}\phi_a^{T_2-T_1}\left|\frac{y_{T_1}^2}{T\sigma^2}\right|^{1/2},
    \label{app:emergence:t2T1T2}
\end{equation}
and thus,
\[
\sum_{T_2=T_1+\lfloor T_{UB}\epsilon\rfloor}^{T_c}t_{T_1,T_2}
\sim_a
\sqrt{\frac{T\rho_a}{2}}\left|\frac{y_{T_1}^2}{T\sigma^2}\right|^{1/2}\sum_{T_2=T_1+\lfloor T_{UB}\epsilon\rfloor}^{T_c}\phi_a^{T_2-T_1}
\sim_a
\sqrt{\frac{T}{2\rho_a}}\phi_a^{(T_c-T_1)}\left|\frac{y_{T_1}^2}{T\sigma^2}\right|^{1/2}.
\]
Then, we obtain
\[
\mbox{EM}_{a,12}^e\sim_a \left|\frac{y_{T_1}^2}{T\sigma^2}\right|^{1/2}\Rightarrow |W(\lambda_1)|.
\]
Similarly, using \eqref{app:emergence:t2T1T2}, we observe that
\[
\mbox{EM}_{b,12}^e\Rightarrow |W(\lambda_1)|.
\]

Next, we derive the asymptotic distributions of $\mbox{LR}_{b,21}^e$ and $\mbox{EM}_{\cdot,21}^e$ for $T_1 > T_2$. Under the null hypothesis of $T_e=T_1$, the standard unit root asymptotic theory implies that
\begin{equation}
    \frac{1}{T_{UB}}\sum_{t=T_2+1}^{T_1}y_{t-1}\Delta y_t\Rightarrow\frac{\sigma^2}{2}\left\{W^2(\lambda_1^*)-W^2(\lambda_2^*)-(\lambda_1^*-\lambda_2^*)\right\}
\label{app:emergence:sum_ydy:T2T1}
\end{equation}
and
\begin{equation}
\frac{1}{T_{UB}^2}\sum_{t=T_2+1}^{T_1}y_{t-1}^2\Rightarrow \sigma^2\int_{\lambda_2^*}^{\lambda_1^*}W^2(s^*)ds^*,
\label{app:emergence:sum_y2:T2T1}
\end{equation}
where $\lambda_j^*\coloneqq T_j/T_{UB}$ for $j=1$ and 2.
From \eqref{app:emergence:sum_ydy:T2T1}, \eqref{app:emergence:sum_y2:T2T1}, and the continuous mapping theorem (CMT), we have
\begin{equation*}
    \mbox{LR}_{a,21}^e
     \sim_a
    \frac{\max_{T_2\in \Lambda_{21}^e}-\sum_{t=T_2+1}^{T_1}y_{t-1}^2}{T_{UB}^2\sigma^2} 
     =
    \frac{-\sum_{t=T_1-\lfloor T\epsilon\rfloor}^{T_1}y_{t-1}^2}{T_{UB}^2\sigma^2}
     \Rightarrow
    -\int_{\lambda_1^*-\epsilon}^{\lambda_1^*}W^2(s^*)ds^*.
\end{equation*}

Using \eqref{app:emergence:sum_ydy:T2T1} and \eqref{app:emergence:sum_y2:T2T1}, we have
\[
t_{T_2,T_1}\Rightarrow \frac{\frac{1}{2}\left[W^2(\lambda_1^*)-W^2(\lambda_2^*)-(\lambda_1^*-\lambda_2^*)\right]}{\sqrt{\int_{\lambda_2^*}^{\lambda_1^*}W^2(s^*)ds^*}}
\eqqcolon ADF(\lambda_2^*,\lambda_1^*).
\]
By the CMT, the above result yields
\[
\mbox{EM}_{a,21}^e\Rightarrow \int_0^{\lambda_1^*-\epsilon}ADF(\lambda_2^*,\lambda_1^*)d\lambda_2^*
\]
and
\[
\mbox{EM}_{b,21}^e\Rightarrow \sup_{0\leq \lambda_2^*\leq \lambda_1^*-\epsilon}ADF(\lambda_2^*,\lambda_1^*).\blacksquare
\]

\noindent
\textbf{Proof of Theorem \ref{theorem:emergence:H1}}: (i) Suppose that $T_e > T_1$ ($\lambda_e > \lambda_1$) and let $T_2 \geq T_e$ to investigate $\mbox{LR}_{\cdot,12}^e$ and $\mbox{EM}_{\cdot,12}^e$. Then, similarly to \eqref{app:emergence:sum_y2:T1T2} and \eqref{app:emergence:sum_y2:T2T1},  we have
\begin{align}
    \sum_{t=T_1+1}^{T_2}y_{t-1}^2
    & = 
        \sum_{t=T_1+1}^{T_e}y_{t-1}^2+\sum_{t=T_e+1}^{T_2}y_{t-1}^2 \nonumber \\
    & = 
    O_p(T^2)+\sum_{t=T_e+1}^{T_2}y_{t-1}^2 \nonumber \\
    & \sim_a
    \frac{\sigma^2}{2}\frac{T\phi_a^{2(T_2-T_e)}}{\rho_a}\left(\frac{y_{T_e^2}}{T\sigma^2}\right),
    \label{app:emergence:Sum_y2:T1ltTe}  
\end{align}
and from \eqref{app:emergence:sum_ydy:T1T2} and \eqref{app:emergence:sum_ydy:T2T1},
\begin{align*}
    \sum_{t=T_1+1}^{T_2}y_{t-1}\Delta y_t
    & = 
    \sum_{t=T_1+1}^{T_e}y_{t-1}\Delta y_t+\sum_{t=T_e+1}^{T_2}y_{t-1}\Delta y_t \nonumber \\
    & =
    O_p(T)+\sum_{t=T_e+1}^{T_2}y_{t-1}\Delta y_t \nonumber \\
    & \sim_a
    \frac{\sigma^2}{2}T\phi_a^{2(T_2-T_e)}\left(\frac{y_{T_e}^2}{T\sigma^2}\right).
\end{align*}
Then, we have
\[
\mbox{LR}_{a,12}^e \leq O_p(\phi_a^{-2(T_e-T_1)})=o_p(1).
\]
$\mbox{LR}_{b,12}^e=o_p(1)$ is proved similarly.

Regarding $\mbox{EM}_{\cdot 12}^e$, noting that
\begin{equation}
t_{T_1,T_2} \sim_a \sqrt{\frac{T\rho_a}{2}}\phi_a^{T_2-T_e}\left|\frac{y_{T_e}^2}{T\sigma^2}\right|^{1/2},
\label{app:emergence:t2:T1TeT2}
\end{equation}
we have
\[
\mbox{EM}_{b,12}^e\sim_a \phi_a^{-(T_e-T_1)}\left|\frac{y_{T_e}^2}{T\sigma^2}\right|^{1/2}=o_p(1).
\]
For $\mbox{EM}_{a,12}^e$, because \eqref{app:emergence:t2:T1TeT2} holds for $T_2\geq T_e$, whereas it is not difficult to see that $t_{T_1,T_2}=O_p(1)$ for $T_2 < T_e$. Using these results, it can be shown that
\[
\mbox{EM}_{a,12}^e\sim_a \frac{\displaystyle\sqrt{\frac{T}{2\rho_a}}\phi_a^{T_c-T_e}\left|\frac{y_e^2}{T\sigma^2}\right|^{1/2}}{\displaystyle \sqrt{\frac{T\phi_a^{2(T_c-T_1)}}{2\rho_a}}}=O_p(\phi_a^{-(T_e-T_1)})=o_p(1).
\]

For $\mbox{LR}_{\cdot,21}^e$ and $\mbox{EM}_{\cdot,21}^e$, because $T_2<T_1 < T_e$, it holds that
\[
    \sum_{t=T_2+1}^{T_1}y_{t-1}^2=O_p(T^2)
    \quad\mbox{and}\quad
    \sum_{t=T_2+1}^{T_1}y_{t-1}\Delta y_t=O_p(T),
\]
and thus,
\[
\mbox{LR}_{a,21}^e=-O_p(1),
\]
where the $O_p(1)$ term is positive, and $t_{T_2,T_1}=O_p(1)$, which implies that $\mbox{EM}_{a,21}$ and $\mbox{EM}_{b,21}^e$ are $O_p(1)$.

\noindent
(ii) Suppose that $T_e < T_1$ ($\lambda_e < \lambda_1$). For $\mbox{LR}_{\cdot, 12}$ and $\mbox{EM}_{\cdot,12}$ with $T_2$ set to be greater than $T_1$, we have,
\[
\sum_{t=T_1+1}^{T_2}y_{t-1}^2\sim_a \frac{\sigma^2}{2}\frac{T\phi_a^{2(T_2-T_e)}}{\rho_a}\left(\frac{y_{T_e}^2}{T\sigma^2}\right)
\quad\mbox{and}\quad
\sum_{t=T_1+1}^{T_2}y_{t-1}\Delta y_t=\frac{\sigma^2}{2}T\phi_a^{2(T_2-T_e)}\left(\frac{y_{T_e}^2}{T\sigma^2}\right)
\]
and then,
\[
\mbox{LR}_{a,12}\sim_a \phi_a^{2(T_1-T_e)}\left(\frac{y_{T_e}^2}{T\sigma^2}\right)\to \infty.
\]
$\mbox{LR}_{b,12}\to \infty$ is proved in exactly the same manner. In addition, we can observe that
\begin{equation}
t_{T_1,T_2}\sim_a \sqrt{\frac{T\rho_a}{2}}\phi_a^{T_2-T_e}\left|\frac{y_{T_e}^2}{T\sigma^2}\right|^{1/2}
\label{app:emergence:t2:TeT1}
\end{equation}
and thus,
\[
\mbox{EM}_{a,12}\sim_a \phi_a^{T_1-T_e}\left|\frac{y_{T_e}^2}{T\sigma^2}\right|^{1/2}\to\infty.
\]
Similarly, we can also show that $\mbox{EM}_{b,12}\to \infty$ from \eqref{app:emergence:t2:TeT1}.

For $\mbox{LR}_{\cdot,21}$ and $\mbox{EM}_{\cdot,21}$, let $T_2 < T_e < T_1$. Then, we have
\[
\sum_{t=T_2+1}^{T_1}y_{t-1}^2\sim_a \sum_{t=T_e+1}^{T_1}y_{t-1}^2\sim_a \frac{\sigma^2}{2}\frac{T\phi_a^{2(T_1-T_e)}}{\rho_a}\left(\frac{y_{T_e}^2}{T\sigma^2}\right)
\]
and
\[
\sum_{t=T_2+1}^{T_1}y_{t-1}\Delta y_t\sim_a \frac{y_{T_1}^2}{2}\sim_a \frac{\sigma^2}{2}T\phi_a^{2(T_1-T_e)}\left(\frac{y_{T_e}^2}{T\sigma^2}\right)
\]
and thus, $\mbox{LR}_{a,21}\to \infty$. Using the above results, it is observed that
\[
t_{T_2,T_1}\sim_a \sqrt{\frac{T\rho_a}{2}}\phi_a^{T_1-T_e}\left|\frac{y_{T_e}^2}{T\sigma^2}\right|^{1/2}
\]
and then, it is shown that $\mbox{EM}_{a,21}$ and $\mbox{EM}_{b,21}$ diverge to infinity.$\blacksquare$

\noindent
\textbf{Proof of Theorem \ref{theorem:collapse:H0}}: We first note that, for $T_1=T_c < t \leq T_r$,
\begin{equation}
    y_t=\phi_b^{(t-T_c)}y_{T_c}+\sum_{j=T_c+1}^{t}\phi_b^{(t-j)}\varepsilon_t.
    \label{app:collapse:y}
\end{equation}
Then, similarly to Lemma 3 of \citet{KurozumiS2023}, we have, noting that $y_{T_c}=y_{T_1}\sim_a \phi_a^{T_1-T_e}y_{T_e}$ under the null hypothesis,
\begin{align}
    \sum_{t=T_1+1}^{T_2}y_{t-1}^2
    & \sim_a
    y_{T_c}^2\sum_{t=T_1+1}^{T_2}\phi_b^{2(t-T_c-1)} \nonumber \\
    & \sim_a
    \frac{T\phi_a^{2(T_1-T_e)}}{2\rho_b}\left(\frac{1}{T}y_{T_e}^2\right),
    \label{app:collapse:sum_y2:T1T2} \\
    \sum_{t=T_1+1}^{T_2}y_{t-1}\varepsilon_t 
    & =
    O_p\left(T^{(\beta+1)/2}\phi_a^{(T_1-T_e)}\right)
    \label{app:collapse:sum_yeps:T1T2} \\
    \sum_{t=T_1+1}^{T_2}y_{t-1}\Delta y_t
    & = 
    \frac{1}{2}\left(y_{T_2}^2-y_{T_1}^2-\sum_{t=T_1+1}^{T_2}(\Delta y_t)^2\right) \nonumber \\
    & \sim_a
    -\frac{1}{2}y_{T_1}^2 \nonumber \\
    & \sim_a
    -\frac{1}{2}T\phi_a^{2(T_1-T_e)}\left(\frac{1}{T}y_{T_e}^2\right)
    \label{app:collapse:sum_ydy:T1T2}.
\end{align}

To derive the limiting distribution of $\mbox{LR}_{a,12}^c$, we transform expression \eqref{LR12:org:c} as
\begin{align*}
    \MoveEqLeft 2\sum_{t=T_1+1}^{T_2}y_{t-1}(y_t-\hat{\phi}_ay_{t-1})+(\hat{\phi}_a-\hat{\phi}_b)\sum_{t=T_1+1}^{T_2}y_{t-1}^2 \\
    & =
    2\sum_{t=T_1+1}^{T_2}y_{t-1}[(\phi_b-\hat{\phi}_b+\hat{\phi}_b)y_{t-1}+\varepsilon_t-\hat{\phi}_ay_{t-1}]+(\hat{\phi}_a-\hat{\phi}_b)\sum_{t=T_1+1}^{T_2}y_{t-1}^2 \\
    & \sim_a
    -(\hat{\phi}_a-\hat{\phi}_b)\sum_{t=T_1+1}^{T_2}y_{t-1}^2 \\
    & \sim_a
    -({\phi}_a-{\phi}_b)\frac{T\phi_a^{2(T_1-T_e)}}{2\rho_b}\left(\frac{1}{T}y_{T_e}^2\right),
\end{align*}
where we used \eqref{app:collapse:sum_y2:T1T2} and \eqref{app:collapse:sum_yeps:T1T2}. As this relation holds uniformly over $T_2\in \Lambda_{12}^c$, we obtain the limiting distribution of $LR_{a,12}^c$.

Using \eqref{app:collapse:sum_y2:T1T2} and \eqref{app:collapse:sum_ydy:T1T2}, we also have
\begin{align*}
    t_{T_1,T_2}
    & =
    \frac{\sum_{t=T_1+1}^{T_2}y_{t-1}\Delta y_t}{\hat{\sigma}\sqrt{\sum_{t=T_1+1}^{T_2}y_{t-1}^2}} \\
    & \sim_a
    \frac{\displaystyle -\frac{1}{2}T\phi_a^{2(T_1-T_e)}\left(\frac{1}{T}y_{T_e}^2\right)}{\displaystyle \sigma\sqrt{\frac{T\phi_a^{2(T_1-T_e)}}{2\rho_b}\left(\frac{1}{T}y_{T_e}^2\right)}} \\
    & \sim_a
    -\sqrt{\frac{T\rho_b }{2}}\phi_a^{T_1-T_e}\left(\frac{1}{T\sigma^2}y_{T_e}^2\right)^{1/2}.
\end{align*}
As the last relation holds uniformly over $T_2\in \Lambda_{12}^c$, we obtain the limiting distributions of $\mbox{EM}_{a,12}^c$ and $\mbox{EM}_{b,12}$.

Next, for $T_e+1\leq t < T_1=T_c$, similarly to \eqref{app:emergence:sum_y2:T1T2}, \eqref{app:emergence:sum_ydy:T1T2}, and Lemma 2 of \citet{KurozumiS2023}, we have, using \eqref{app:emergence:y},
\begin{align}
    \sum_{t=T_2+1}^{T_1}y_{t-1}^2
    & \sim_a
    y_{T_e}^2\sum_{t=T_2+1}^{T_1}\phi_a^{2(t-1-T_e)} \nonumber \\
    & \sim_a
    \frac{T\phi_a^{2(T_1-T_e)}}{2\rho_a}\left(\frac{1}{T}y_{T_e}^2\right),
    \label{app:collapse:sum_y2:T2T1} \\
    \sum_{t=T_2+1}^{T_1}y_{t-1}\varepsilon_t 
    & =
    O_p\left(T^{(\alpha+1)/2}\phi_a^{T_1-T_e}\right),
    \label{app:collapse:sum_yeps:T2T1} \\
    \sum_{t=T_2+1}^{T_1}y_{t-1}\Delta y_t
    & = 
    \frac{1}{2}\left(y_{T_1}^2-y_{T_2}^2-\sum_{t=T_2+1}^{T_1}(\Delta y_t)^2\right) \nonumber \\
    & \sim_a
    \frac{1}{2}y_{T_1}^2 \nonumber \\
    & \sim_a
    \frac{1}{2}T\phi_a^{2(T_1-T_e)}\left(\frac{1}{T}y_{T_e}^2\right)
    \label{app:collapse:sum_ydy:T2T1}.
\end{align}

To derive the limiting distribution of $\mbox{LR}_{a,21}^c$, we transform expression \eqref{LR21:org:c} as
\begin{align*}
    \MoveEqLeft 2\sum_{t=T_2+1}^{T_1}y_{t-1}(y_t-\hat{\phi}_ay_{t-1})+(\hat{\phi}_a-\hat{\phi}_b)\sum_{t=T_2+1}^{T_1}y_{t-1}^2 \\
    & =
    2\sum_{t=T_2+1}^{T_1}y_{t-1}[(\phi_a-\hat{\phi}_a)y_{t-1}+\varepsilon_t]+(\hat{\phi}_a-\hat{\phi}_b)\sum_{t=T_2+1}^{T_1}y_{t-1}^2 \\
    & \sim_a
    (\hat{\phi}_a-\hat{\phi}_b)\sum_{t=T_2+1}^{T_1}y_{t-1}^2 \\
    & \sim_a
    ({\phi}_a-{\phi}_b)\frac{T\phi_a^{2(T_1-T_e)}}{2\rho_a}\left(\frac{1}{T}y_{T_e}^2\right),
\end{align*}
where we used \eqref{app:collapse:sum_y2:T2T1} and \eqref{app:collapse:sum_yeps:T2T1}. As this relation holds uniformly over $T_2\in \Lambda_{21}^c$, we obtain the limiting distribution of $LR_{a,21}^c$.

Using \eqref{app:collapse:sum_y2:T2T1} and \eqref{app:collapse:sum_ydy:T2T1}, we also have
\begin{align*}
    t_{T_2,T_1}
    & =
    \frac{\sum_{t=T_2+1}^{T_1}y_{t-1}\Delta y_t}{\hat{\sigma}\sqrt{\sum_{t=T_2+1}^{T_1}y_{t-1}^2}} \\
    & \sim_a
    \frac{\displaystyle \frac{1}{2}T\phi_a^{2(T_1-T_e)}\left(\frac{1}{T}y_{T_e}^2\right)}{\displaystyle \sigma\sqrt{\frac{T\phi_a^{2(T_1-T_e)}}{2\rho_a}\left(\frac{1}{T}y_{T_e}^2\right)}} \\
    & \sim_a
    \sqrt{\frac{1}{2}T\rho_a}\phi_a^{T_1-T_e}\left(\frac{1}{T\sigma^2}y_{T_e}^2\right)^{1/2}.
\end{align*}
As the last relation holds uniformly over $T_2\in \Lambda_{21}^c$, we obtain the limiting distributions of $\mbox{EM}_{a,21}^c$ and $\mbox{EM}_{b,21}^c$.$\blacksquare$

\noindent
\textbf{Proof of Theorem \ref{theorem:collapse:H1}}: (i) Suppose that $T_c > T_1$ ($\lambda_c > \lambda_1$) and let $T_2= T_c$ to investigate $\mbox{LR}_{a,12}$ and $\mbox{EM}_{\cdot,12}$. Similarly to \eqref{app:emergence:sum_y2:T1T2} and \eqref{app:emergence:sum_ydy:T1T2}, from \eqref{app:emergence:y}, we have
\begin{equation}
    \sum_{t=T_1+1}^{T_2}y_{t-1}^2
    \sim_a
    \phi_a^{2(T_c-T_e)}\frac{y_{T_e}^2}{2\rho_a},
    \label{app:collapse:sum_y2:T1TcT2}
\end{equation}
and 
\begin{equation}
    \sum_{t=T_1+1}^{T_2}y_{t-1}\Delta y_t
    \sim_a
    \frac{1}{2}y_{T_2}^2
    \sim_a
    \frac{1}{2}\phi_a^{2(T_c-T_e)}y_{T_e}^2.
    \label{app:collapse:sum_ydy:T1TcT2}
\end{equation}
\eqref{app:collapse:sum_y2:T1TcT2} and \eqref{app:collapse:sum_ydy:T1TcT2} yield
\begin{align*}
    \mbox{LR}_{a,12}^c 
    & \geq
    \frac{2\sum_{t=T_1+1}^{T_2}y_{t-1}\Delta y_t+(2-\phi_a-\phi_b)\sum_{t=T_1+1}^{T2}y_{t-1}^2}{\sigma^2T(\phi_a-\phi_b)\phi_a^{2(T_1-T_e)}/(2\rho_a)} \\
    & \sim_a
    \frac{\left[\phi_a^{2(T_c-T_e)}+(-\frac{a}{T^{\alpha}}+\frac{b}{T^{\beta}})\frac{T^{\alpha}}{2a}\phi_a^{2(T_c-T_e)}\right]y_{T_e}^2}{\sigma^2T(\frac{a}{T^{\alpha}}+\frac{b}{T^{\beta}})\phi_a^{2(T_1-T_e)}\frac{T^{\alpha}}{2a}} \\
    & \geq
    O_p\left(\frac{\phi_a^{2(T_c-T_1)}}{1+\frac{b}{a}T^{\alpha-\beta}}\right) \\
    & \to \infty.
\end{align*}
As
\[
    t_{T_1,T_2}
    \sim_a
    \frac{\frac{1}{2}\phi_a^{2(T_c-T_e)}y_{T_e}^2}{\sigma \sqrt{\frac{T^{\alpha}}{2a}\phi_a^{2(T_c-T_e)}y_{T_e}^2}}
    =
    O_p(T^{(1-\alpha)/2}\phi_a^{T_c-T_e})
\]
uniformly over $T_2\in\Lambda_{12}^c$, we can observe that
\[
    \mbox{EM}_{a,12}^c,\;\mbox{EM}_{b,12}^c
    =
    O_p\left(T^{(\beta-\alpha)/2}\phi_a^{T_c-T_1}\right)
    \to
    \infty.
\]

For $T_2 < T_1$, we have
\begin{equation}
    \sum_{t=T_2+1}^{T_1}y_{t-1}^2
    \sim_a
    \frac{1}{2\rho_a}\phi_a^{2(T_1-T_e)}y_{T_e}^2,
    \label{app:collapose:sum_y2:T2T1Tc}
\end{equation}
and
\begin{equation}
    \sum_{t=T_2+1}^{T_1}y_{t-1}\Delta y_t
    \sim_a
    \frac{1}{2}y_{T_1}^2
    \sim_a
    \frac{1}{2}\phi_a^{2(T_1-T_e)}y_{T_e}^2.
    \label{app:collapse:sum_ydy:T2T1Tc}
\end{equation}
Using \eqref{app:collapose:sum_y2:T2T1Tc} and \eqref{app:collapse:sum_ydy:T2T1Tc}, we have
\begin{align*}
    \MoveEqLeft 2\sum_{t=T_2+1}^{T_1}y_{t-1}\Delta y_t+(2-\hat{\phi}_a-\hat{\phi}_b)\sum_{t=T_2+1}^{T_1}y_{t-1}^2 \\
    & \sim_a
    \phi_a^{2(T_1-T_e)}y_{T_e}^2+\left(-\frac{a}{T^{\alpha}}+\frac{b}{T^{\beta}}\right)\frac{T^{\alpha}}{2a}\phi_a^{2(T_1-T_e)}y_{T_e}^2 \\
    & = 
    \frac{1}{2}\phi_a^{2(T_1-T_e)}y_{T_e}^2+\frac{b}{2a}T^{\alpha-\beta}\phi_a^{2(T_1-T_e)}y_{T_e}^2.
\end{align*}
Noting that the last expression holds uniformly over $T_2\in\Lambda_{21}^c$, we conclude that 
\[
    \mbox{LR}_{a,21}
    = 
    O_p\left(\frac{T\phi_a^{2(T_1-T_e)}+T^{1+\alpha-\beta}\phi_a^{2(T_1-T_e)}}{T(\frac{a}{T^{\alpha}}+\frac{b}{T^{\beta}})\phi_a^{2(T_1-T_e)}\frac{T^{\alpha}}{2a}}\right)
    =
    O_p(1).
\]
Similarly, we observe that
\[
    t_{T_2,T_1}
    =
    \frac{O_p\left(T\phi_a^{2(T_1-T_e)}\right)}{T^{(1+\alpha)/2}\phi_a^{(T_1-T_e)}}
    =
    O_p\left(T^{(1-\alpha)/2}\phi_a^{T_1-T_e}\right).
\]
and thus, we have
\[
    \mbox{EM}_{a,21},\;\mbox{EM}_{b,21}
    =
    O_p\left(\frac{T^{(1-\alpha)/2}\phi_a^{T_1-T_e}}{T^{(1-\alpha)/2}\phi_a^{T_1-T_e}}\right)=O_p(1).
\]

\noindent
(ii) We prove only the case where $\alpha < \beta$, in which the exploding speed is faster than the collapsing speed. The proof of the case where $\alpha  \geq \beta$ proceeds in exactly the same manner.

Suppose that $T_c < T_1$ ($\lambda_c < \lambda_1$) and let $T_2 > T_1$. In exactly the same manner as \eqref{app:collapse:sum_y2:T1T2}, \eqref{app:collapse:sum_ydy:T1T2}, and Lemma 3 of \citet{KurozumiS2023}, using \eqref{app:collapse:y}, we observe that
\begin{align}
    \sum_{t=T_1+1}^{T_2}y_{t-1}^2
    & \sim_a
    y_{T_c}^2\sum_{t=T_1+1}^{T_2}\phi_b^{2(t-T_c-1)} \nonumber \\
    & =
    O_p\left(T^{\beta+1}\phi_b^{2(T_1-T_c)}\phi_a^{2(T_c-T_e)}\right),
    \label{app:collapse:sum_y2:TcT1T2}
\end{align}
while
\begin{align*}
    y_{T_1}^2
    & \sim_a
    \phi_b^{2(T_1-T_c)}\phi_a^{2(T_c-T_e)}y_{T_e}^2=
    O_p\left(T\phi_b^{2(T_1-T_c)}\phi_a^{2(T_c-T_e)}\right), \\
    y_{T_2}^2
    & =
    O_p(T\phi_b^{2(T_2-T_c)}\phi_a^{2(T_c-T_e)}), \\
    \sum_{t=T_1+1}^{T_2}(\Delta y_{t})^2
    & \sim_a
    (\phi_b-1)^2\sum_{t=T_1+1}^{T_2}y_{t-1}^2=    O_p\left(T^{1-\beta}\phi_b^{2(T_1-T_c)}\phi_a^{2(T_c-T_e)}\right),
\end{align*}
which implies
\begin{equation}
    \sum_{t=T_1+1}^{T_2}y_{t-1}\Delta y_t
    =
    O_p\left(T\phi_b^{2(T_1-T_c)}\phi_a^{2(T_c-T_e)}\right).
    \label{app:collapose:sum_ydy:TcT1T2}
\end{equation}
From \eqref{app:collapse:sum_y2:TcT1T2} and \eqref{app:collapose:sum_ydy:TcT1T2}, we obtain, as $\phi_a^{2(T_c-T_1)}\to 0$ at an exponential rate,
\begin{align*}
    \MoveEqLeft \mbox{LR}_{a,12}^c \\
    & =
    \frac{O_p\left(T\phi_b^{2(T_1-T_c)}\phi_a^{2(T_c-T_e)}\right)+O_p(T^{-\alpha}+T^{-\beta})O_p\left(T^{\beta+1}\phi_b^{2(T_1-T_c)}\phi_a^{2(T_c-T_e)}\right)}{O_p\left(T(T^{-\alpha}+T^{-\beta})\phi_a^{2(T_1-T_e)}T^{\beta}\right)} \\
    & =
    o_p(1).
\end{align*}
Similarly, observing that
\begin{align*}
    t_{T_1,T_2}
    & =
    \frac{O_p\left(T\phi_b^{2(T_1-T_c)}\phi_a^{2(T_c-T_e)}\right)}{O_p\left(T^{(\beta+1)/2}\phi_b^{T_1-T_c}\phi_a^{T_c-T_e}\right)}=O_p\left(T^{(1-\beta)/2}\phi_b^{T_1-T_c}\phi_a^{T_c-T_e}\right),
\end{align*}
we can show that $\mbox{EM}_{a,12}^c$ and $\mbox{EM}_{a,21}^c$ are $o_p(1)$.

For $\mbox{LR}_{a,21}^c$ and $\mbox{EM}_{\cdot,21}^c$, consider the case where $T_2=T_c$. Then, we observe using \eqref{app:collapse:y} that
\begin{equation}
    \sum_{t=T_2+1}^{T_1}y_{t-1}^2
    \sim_a
    y_{T_c}^2\sum_{t=T_c+1}^{T_1}\phi_b^{2(t-T_c-1)}\sim_a\frac{T^{\beta}}{2b}\phi_a^{2(T_c-T_e)}y_{T_e}^2,
    \label{app:collapse:sum_y2:T2TcT1}
\end{equation}
\begin{equation}
    \sum_{t=T_2+1}^{T_1}y_{t-1}\Delta y_t
    \sim_a
    -\frac{1}{2}y_{T_c}^2
    \sim_a
    -\frac{1}{2}\phi_a^{2(T_c-T_e)}y_{T_e}^2,
    \label{app:collapse:sum_ydy:T2TcT1}
\end{equation}
from which
\begin{align*}
    \MoveEqLeft 2\sum_{t=T_2+1}^{T_1}y_{t-1}\Delta y_t+(2-\phi_a-\phi_b)\sum_{t=T_2+1}^{T_1}y_{t-1}^2 \\
    & \sim_a
    -\phi_a^{2(T_c-T_e)}y_{T_e}^2+\left(-\frac{a}{T^{\alpha}}+\frac{b}{T^{\beta}}\right)\frac{T^{\beta}}{2b}\phi_a^{2(T_c-T_e)}y_{T_e}^2 \\
    & =
    O_p\left(T\phi_a^{2(T_c-T_e)}\right)+O_p\left(T^{1+\beta-\alpha}\phi_a^{2(T_c-T_e)}\right).
\end{align*}
Therefore, $\mbox{LR}_{a,21}^c$ is shown to be $o_p(1)$.

Similarly, using \eqref{app:collapse:sum_y2:T2TcT1} and \eqref{app:collapse:sum_ydy:T2TcT1}, $\mbox{EM}_{a,21}^c$ and $\mbox{EM}_{b,21}^c$ are proved to be $o_p(1)$.$\blacksquare$

\noindent
\textbf{Proof of Theorem \ref{theorem:recovery:H0}}: (i) When $\alpha < \beta$, we can see that $\phi_a^{T_c-T_e}\phi_b^{T_r-T_c}$ diverges to infinity at the exponential rate. Then, for $T_1=T_r < t \leq T$,
\begin{equation}
    y_t=y_{T_1}+\sum_{j=T_1+1}^{t}\varepsilon_t\sim_a \phi_a^{T_c-T_e}\phi_b^{T_1-T_c}y_{T_e}
    \label{app:recovery:y:ab}
\end{equation}
holds uniformly under the null hypothesis. Using \eqref{app:recovery:y:ab}, we have
\begin{equation}
    \sum_{t=T_1+1}^{T_2}y_{t-1}^2\sim_a (T_2-T_1)\phi_a^{2(T_c-T_e)}\phi_b^{2(T_1-T_c)}y_{T_e}^2,
    \label{app:recovery:sum_y2:T1T2:ab}
\end{equation}
\begin{align}
    \sum_{t=T_1+1}^{T_2}y_{t-1}\Delta y_t
    & = 
    \frac{1}{2}\left(y_{T_2}^2-y_{T_1}^2-\sum_{t=T_1+1}^{T_2}(\Delta y_t)^2\right) \nonumber \\
    & \sim_a
    y_{T_1}\sum_{t=T_1+1}^{T_2}\varepsilon_t \nonumber \\
    & \sim_a
    \phi_a^{T_c-T_e}\phi_b^{T_1-T_c}y_{T_e}\sum_{t=T_1+1}^{T_2}\varepsilon_t
    \label{app:recovery:sum_ydy:T1T2:ab}.
\end{align}
Then, using \eqref{app:recovery:sum_y2:T1T2:ab} and \eqref{app:recovery:sum_ydy:T1T2:ab}, we obtain
\begin{align*}
    2\sum_{t=T_1+1}^{T_2}y_{t-1}\Delta y_t+\rho_b\sum_{t=T_1+1}^{T_2}y_{t-1}^2
    & \sim_a
    \rho_b (T_2-T_1)\phi_a^{2(T_c-T_e)}\phi_b^{2(T_1-T_c)}y_{T_e}^2,
\end{align*}
which leads to
\[
\mbox{LR}_{a,12}^r\sim_a\frac{1}{T\sigma^2}y_{T_e}^2\Rightarrow W^2(\lambda_e).
\]
For the EM-type tests, as
\begin{align*}
    t_{T_1,T_2}
    & \sim_a
    \frac{\displaystyle \phi_a^{T_c-T_e}\phi_b^{T_1-T_c}y_{T_e}\sum_{t=T_1+1}^{T_2}\varepsilon_t}{\displaystyle \sigma\sqrt{(T_2-T_1)\phi_a^{2(T_c-T_e)}\phi_b^{2(T_1-T_c)}y_{T_e}^2}} \\
    & =
    \frac{\sqrt{T_{CU}}}{ \sqrt{(T_2-T_c)-(T_1-T_c)}}\frac{1}{\sigma\sqrt{T_{CU}}}\mbox{sign}(y_{T_e})\sum_{j=(T_1-T_c)+1}^{T_2-T_c}\varepsilon_{T_c+j} \\
    & \Rightarrow
    \frac{W(\lambda_2^*)-W(\lambda_1^*)}{\sqrt{\lambda_2^*-\lambda_1^*}},
\end{align*}
where $\lambda_j^*\coloneqq (T_j-T_c)/T_{CU}$, because $y_{T_e}$ is supposed to be positive in our model. Using this result, we have the limiting distributions of $\mbox{EM}_{a,12}^r$ and $\mbox{EM}_{b,12}^r$.

To derive the limiting distributions of $\mbox{LR}_{a,21}^r$ and $\mbox{EM}_{\cdot,21}^r$, we note that for $T_c+1\leq t < T_1=T_r$, when $\alpha < \beta$,
\[
    y_t=\phi_b^{t-T_c}y_{T_c}+\sum_{j=T_c+1}^t\phi_b^{t-j}\varepsilon_j\sim_a \phi_a^{T_c-T_e}\phi_b^{t-T_c}y_{T_e}.
\]
As this relation holds uniformly, we have
\begin{align}
    \sum_{t=T_2+1}^{T_1}y_{t-1}^2
    & \sim_a
    \phi_a^{2(T_c-T_e)}y_{T_e}^2\sum_{t=T_2+1}^{T_1}\phi_b^{2(t-T_c-1)} 
    \sim_a
    \frac{1}{2\rho_b}\phi_a^{2(T_c-T_e)}\phi_b^{2(T_2-T_c)}y_{T_e}^2,
    \label{app:recovery:sum_y2:T2T1:ab}
\end{align}
and, because $y_{T_2}^2$ is shown to dominate $y_{T_1}^2$ and $\sum_{t=T_2+1}^{T_1}(\Delta y_t)^2,$
\begin{align}
    \sum_{t=T_2+1}^{T_1}y_{t-1}\Delta y_t
    & \sim_a
    -\frac{1}{2}y_{T_2}^2 
     \sim_a
    -\frac{1}{2}\phi_a^{2(T_c-T_e)}\phi_b^{2(T_2-T_c)}y_{T_e}^2.
    \label{app:recovery:sum_ydy:T2T1:ab}
\end{align}
Then, using \eqref{app:recovery:sum_y2:T2T1:ab} and \eqref{app:recovery:sum_ydy:T2T1:ab}, we obtain
\[
    2\sum_{t=T_2+1}^{T_1}y_{t-1}\Delta y_t+\rho_b\sum_{t=T_2+1}^{T_1}y_{t-1}^2\sim_a
    -\frac{1}{2}\phi_a^{2(T_c-T_e)}\phi_b^{2(T_2-T_c)}y_{T_e}^2,
\]
which leads to
\[
\mbox{LR}_{a,21}^r \sim_a-\frac{y_{T_e}^2}{T\sigma^2}\Rightarrow -W^2(\lambda_e).
\]
For the EM-type tests, as
\[
    t_{T_2,T_1}\sim_a\frac{-\frac{1}{2}\phi_a^{2(T_c-T_e)}\phi_b^{2(T_2-T_c)}y_{T_e}^2}{\sigma\sqrt{\frac{1}{2\rho_b}\phi_a^{2(T_c-T_e)}\phi_b^{2(T_2-T_c)}y_{T_e}^2}}
    =
    -\sqrt{\frac{\rho_b}{2}}\phi_a^{T_c-T_e}\phi_b^{T_2-T_c}\frac{|y_{T_e}|}{\sigma},
\]
we observe that
\[
\mbox{EM}_{a,21}^r,\;\mbox{EM}_{b,21}^r\sim_a -\frac{|y_{T_e}|}{\sigma\sqrt{T}}\Rightarrow -|W(\lambda_e)|.
\]

\noindent
(ii) When $\alpha > \beta$, we can see that $\phi_a^{T_c-T_e}\phi_b^{T_r-T_c}$ converges to zero at the exponential order. Then, we observe that
\[
y_{T_r}=\phi_b^{T_r-T_c}y_{T_c}+\sum_{t=T_c+1}^{T_r}\phi_b^{(T_r-t)}\varepsilon_t=O_p(T^{\beta/2}).
\]
Then, noting that $\beta/2 < 1/2$ because $0<\beta<1$, for $T_1=T_r < t \leq T$, we have
\begin{equation}
    y_t=y_{T_1}+\sum_{t=T_1+1}^t\varepsilon_t=O_p(T^{\beta/2})+\sum_{t=T_1+1}^t\varepsilon_t.
    \label{app:recovery:y:ba}
\end{equation}
Using \eqref{app:recovery:y:ba}, we have
\begin{equation}
    \sum_{t=T_1+1}^{T_2}y_{t-1}^2
    \sim_a
    \sum_{t=T_1+1}^{T_2}\left(\sum_{j=T_1+1}^{t-1}\varepsilon_j\right)^2,
    \label{app:recovery:sum_y2:T1T2:ba}
\end{equation}
\begin{align}
    \sum_{t=T_1+1}^{T_2}y_{t-1}\Delta y_t
    &\sim_a
    \frac{1}{2}\left[y_{T_2}^2-\sum_{t=T_1+1}^{T_2}(\Delta y_t)^2\right] \nonumber \\
    & \sim_a
    \frac{1}{2}\left[\left(\sum_{t=T_1+1}^{T_2}\varepsilon_t\right)^2-\sum_{t=T_1+1}^{T_2}\varepsilon_t^2\right].
    \label{app:recovery:sum_ydy:T1T2:ba}
\end{align}
Then, using \eqref{app:recovery:sum_y2:T1T2:ba} and \eqref{app:recovery:sum_ydy:T1T2:ba}, we obtain
\[
2\sum_{t=T_1+1}^{T_2}y_{t-1}\Delta y_t+\rho_b\sum_{t=T_1+1}^{T_2}y_{t-1}^2\sim_a
\rho_b\sum_{t=T+1}^{T_2}\left(\sum_{j=T_1+1}^{t-1}\varepsilon_j\right)^2=O_p(T^{2-\beta}),
\]
which implies that $\mbox{LR}_{a,12}^r$ diverges to infinity because the scalar goes to zero much faster than $T^{2-\beta}$.
For the EM-type tests, as
\[
t_{T_1,T_2}
\Rightarrow
\frac{\frac{1}{2}\left[\left(W(\lambda_2^*)-W(\lambda_1^*)\right)^2-(\lambda_2^*-\lambda_1^*)\right]}{\sqrt{\int_{\lambda_1^*}^{\lambda_2^*}\left(W(s^*)-W(\lambda_1^*)\right)^2ds^*}}\eqqcolon ADF^r(\lambda_1^*,\lambda_2^*),
\]
we obtain the limiting distributions of $\mbox{EM}_{a,12}$ and $\mbox{EM}_{b,12}$.

To see the behavior of $\mbox{LR}_{a,21}^r$ and $\mbox{EM}_{\cdot,21}^r$, note that for $T_c+\lfloor T_{CU}\epsilon\rfloor < t \leq T_r$, when $\alpha > \beta$,
\[
    y_t=\phi_b^{t-T_c}y_{T_c}+\sum_{j=T_c+1}^{t}\phi_b^{t-j}\varepsilon_j\sim_a \sum_{j=T_c+1}^{t}\phi_b^{t-j}\varepsilon_j.
\]
Then, we observe that
\begin{equation}
    \frac{1}{T_{CU}T^{\beta}}\sum_{t=T_2+1}^{T_1}y_{t-1}^2\sim_a \frac{1}{T_{CU}T^{\beta}}\sum_{t=T_2+1}^{T_1}\left(\sum_{j=T_c+1}^{t-1}\phi_b^{t-1-j}\varepsilon_j\right)^2\CP (\lambda_1^*-\lambda_2^*)\frac{\sigma^2}{2b},
    \label{app:recovery:sum_y2:T2T1:ba}
\end{equation}
where the convergence in probability holds by Lemma 2(a) of \citet{KH09} by noting that the term in the parentheses corresponds to the mildly integrated process, and the convergence in probability holds uniformly over $\epsilon\leq \lambda_2^*\leq \lambda_1^*-\epsilon$ by Lemma A.10 of \citet{Hansen2000b}. In addition, as $y_{T_1}^2=O_p(T^{\beta})$ and $y_{T_2}^2=O_p(T^{\beta})$ whereas
\[
\sum_{t=T_2+1}^{T_1}\left(\Delta y_t\right)^2
=
\sum_{t=T_2+1}^{T_1}\left(-\rho_b y_{t-1}+\varepsilon_t\right)^2
\sim_a
\sum_{t=T_2+1}^{T_1}\varepsilon_t^2=O_p(T),
\]
we observe that
\begin{equation}
    \frac{1}{T_{CU}}\sum_{t=T_2+1}^{T_1}y_{t-1}\Delta y_t \sim_a 
    -\frac{1}{2T_{CU}}\sum_{t=T_2+1}^{T_1}\varepsilon_t^2\CP -\frac{\sigma^2}{2}(\lambda_1^*-\lambda_2^*).
    \label{app:recovery:sum_ydy:T2T1:ba}
\end{equation}
From \eqref{app:recovery:sum_y2:T2T1:ba} and \eqref{app:recovery:sum_ydy:T2T1:ba}, we have
\begin{align}
    \MoveEqLeft
    \frac{1}{T_{CU}}\left(2\sum_{t=T_2+1}^{T_1}y_{t-1}\Delta y_t+\rho_b\sum_{t=T_2+1}^{T_1}y_{t-1}^2\right) \nonumber \\
    & \sim_a 
    -\frac{1}{T_{CU}}\sum_{T_2+1}^{T_1}\varepsilon_t^2+
    \frac{b}{T^{\beta}T_{CU}}\sum_{t=T_2+1}^{T_1}\left(\sum_{j=T_c+1}^{t-1}\phi_b^{t-1-j}\varepsilon_j\right)^2 \nonumber \\
    & \CP
    -(\lambda_1^*-\lambda_2^*)\sigma^2+b\frac{\sigma^2}{2b}(\lambda_1^*-\lambda_2^*) \nonumber \\
    & =
    -\frac{1}{2}(\lambda_1^*-\lambda_2^*)\sigma^2,
    \label{app:recovery:LR21:num:ba}
\end{align}
which implies that $\mbox{LR}_{a,21}^r \to -\infty$. For the EM-type tests, using \eqref{app:recovery:sum_y2:T2T1:ba} and \eqref{app:recovery:sum_ydy:T2T1:ba}, we observe that
\begin{align*}
    \sqrt{\frac{T^{\beta}}{T_{CU}}}t_{T_2,T_1}
    & \sim_a 
    \frac{-\frac{1}{2T_{CU}}\sum_{t=T_2+1}^{T_1}\varepsilon_t^2}{\sigma\sqrt{\frac{1}{T_{CU}T^{\beta}}\sum_{t=T_2+1}^{T_1}\left(\sum_{j=T_c+1}^{t-j-1}\varepsilon_j\right)^2}} \\
    & \CP
    -\sqrt{\frac{b(\lambda_1^*-\lambda_2^*)}{2}},
\end{align*}
which implies that $\mbox{EM}_{a,21},\;\mbox{EM}_{b,21}\to -\infty$.$\blacksquare$

\noindent
\textbf{Proof of Theorem \ref{theorem:recovery:H1}}: (i-a) When $\alpha < \beta$, suppose that $T_r > T_1$ ($\lambda_r > \lambda_1$). Then, as $y_t\sim_a \phi_a^{T_c-T_e}\phi_b^{t-T_c}y_{T_e}$ for $T_c+1\leq t \leq T_r$, we observe for $T_1 < T_2\leq T_r$ that, similarly to \eqref{app:recovery:sum_y2:T2T1:ab} and \eqref{app:recovery:sum_ydy:T2T1:ab},
\[
\sum_{t=T_1+1}^{T_2}y_{t-1}^2\sim_a \phi_a^{2(T_c-T_e)}y_{T_e}^2\sum_{t=T_1+1}^{T_2}\phi_b^{2(t-1-T_c)}\sim_a\frac{1}{2\rho_b}\phi_a^{2(T_c-T_e)}\phi_b^{2(T_1-T_c)}y_{T_e}^2,
\]
and
\[
\sum_{t=T_1+1}^{T_2}y_{t-1}\Delta y_t\sim_a -\frac{1}{2}y_{T_1}^2\sim_a -\frac{1}{2}\phi_a^{2(T_c-T_e)}\phi_b^{2(T_1-T_c)}y_{T_e}^2.
\]
Using these results, we have
\[
2\sum_{t=T_1+1}^{T_2}y_{t-1}\Delta y_t+\rho_b\sum_{t=T_1+1}^{T_2}y_{t-1}^2\sim_a-\frac{1}{2}\phi_a^{2(T_c-T_e)}\phi_b^{2(T_1-T_c)}y_{T_e}^2,
\]
and
\[
t_{T_1,T_2}\sim_a\frac{-\frac{1}{2}\phi_a^{2(T_c-T_e)}\phi_b^{2(T_1-T_c)}y_{T_e}^2}{\sigma\sqrt{\frac{1}{2\rho_b}\phi_a^{2(T_c-T_e)}\phi_b^{2(T_1-T_c)}y_{T_e}^2}}
\sim_a-\sqrt{\frac{\rho_b}{2}}\phi_a^{T_c-T_e}\phi_b^{T_1-T_c}|y_{T_e}|.
\]
As the same results are obtained for $T_2 > T_r$, we observe from the definitions of the test statistics that
\[
\mbox{LR}_{a,12}^r \to 0\quad\mbox{and}\quad \mbox{EM}_{a,12},\;\mbox{EM}_{b,12}\to -\infty.
\]

The orders of $\mbox{LR}_{\cdot,21}^r$ and $\mbox{EM}_{\cdot,21}^r$ are obtained in exactly the same manner as the derivations under the null hypothesis and we omit the details.

\noindent
(i-b) Suppose that $T_r < T_1$ ($\lambda_r < \lambda_1$). Similarly to \eqref{app:recovery:sum_y2:T1T2:ab} and \eqref{app:recovery:sum_ydy:T1T2:ab}, we observe that
\[
\sum_{t=T_1+1}^{T_2}y_{t-1}^2\sim_a (T_2-T_1)\phi_a^{2(T_c-T_e)}\phi_b^{2(T_r-T_c)}y_{T_e}^2
\]
and
\[
\sum_{t=T_1+1}^{T_2}y_{t-1}\Delta y_t
\sim_a
\phi_a^{T_c-T_e}\phi_b^{T_r-T_c}y_{T_e}\sum_{t=T_r+1}^{T_2}\varepsilon_t.
\]
Then, we have
\[
2\sum_{T_1+1}^{T_2}y_{t-1}\Delta y_t+\rho_b\sum_{t=T_1+1}^{T_2}y_{t-1}^2\sim_a (T_2-T_1)\rho_b\phi_a^{2(T_c-T_e)}\phi_b^{2(T_r-T_c)}y_{T_e}^2,
\]
which implies $\mbox{LR}_{a,12}\to \infty$. In addition, we can observe that $t_{T_1,T_2}\sim_a O_p(1)$ uniformly over $T_2\in \Lambda_{12}^r$ and thus, $\mbox{EM}_{a,12}^r$ and $\mbox{EM}_{b,12}^r$ are $O_p(1)$.

For $T_2 < T_r$, we observe that
\begin{align*}
    \sum_{t=T_2+1}^{T_1}y_{t-1}^2
    & = 
    \sum_{t=T_2+1}^{T_r}y_{t-1}^2+
    \sum_{t=T_r+1}^{T_1}y_{t-1}^2 \\
    & \sim_a
    \sum_{t=T_2+1}^{T_r}y_{t-1}^2 \\
    & \sim_a
    \frac{1}{2\rho_b}\phi_a^{2(T_c-T_e)}\phi_b^{2(T_2-T_c)}y_{T_e}^2,
\end{align*}
and
\begin{align*}
    \sum_{t=T_2+1}^{T_1}y_{t-1}\Delta y_t
    & \sim_a
    -\frac{1}{2}y_{T_2}^2 \\
    & \sim_a
    -\frac{1}{2}\phi_a^{2(T_c-T_e)}\phi_b^{2(T_2-T_c)}y_{T_e}^2,
\end{align*}
which imply that
\[
2\sum_{t=T_2+1}^{T_1}y_{t-1}\Delta y_t+\rho_b\sum_{t=T_2+1}^{T_1}y_{t-1}^2\sim_a -\frac{1}{2}\phi_a^{2(T_c-T_r)}\phi_b^{2(T_2-T_c)}y_{T_e}^2,
\]
whereas for $T_r \leq T_2 < T_1$,
\begin{align*}
    \sum_{t=T_2+1}^{T_1}y_{t-1}^2
    & \sim_a
    (T_1-T_2)\phi_a^{2(T_c-T_e)}\phi_b^{2(T_r-T_c)}y_{T_e}^2,
\end{align*}
and 
\[
\sum_{t=T_2+1}^{T_1}y_{t-1}\Delta y_t\sim_a \phi_a^{T_c-T_e}\phi_b^{T_r-T_c}y_{T_e}\sum_{t=T_r+1}^{T_1}\varepsilon_t,
\]
which imply that
\[
2\sum_{t=T_2+1}^{T_1}y_{t-1}\Delta y_t+\rho_b\sum_{t=T_2+1}^{T_1}y_{t-1}^2\sim_a (T_1-T_2)\rho_b\phi_a^{2(T_c-T_e)}\phi_b^{2(T_r-T_c)}y_{T_e}^2.
\]
From these results, we observe that the numerator of $\mbox{LR}_{a,21}$ cannot be maximized for $T_2 < T_r$. Therefore, from the definition of the test statistic, we observe that $\mbox{LR}_{a,21} \to \infty$. In addition, we observe that
\[
t_{T_2,T_1}\Rightarrow \frac{W(\lambda_1^*)-W(\lambda_r^*)}{\sqrt{\lambda_1^*-\lambda_2^*}}
\]
for $T_2 \geq T_r$, whereas for $T_2 < T_r$,
\[
t_{T_2,T_1}\sim_a -\sqrt{\frac{\rho_b}{2}}\phi_a^{T_c-T_e}\phi_b^{T_2-T_c}\frac{|y_{T_e}|}{\sigma}.
\]
These imply that $\mbox{EM}_{a,21}=O_p(1)$ while $\mbox{EM}_{b,21}=o_p(1)$.

\noindent
(ii-a) When $\alpha > \beta$, suppose that $T_r > T_1$ ($\lambda_r > \lambda_1$). Then, following \eqref{app:recovery:LR21:num:ba}, we observe that, for $T_2=T_r$,
\[
\frac{1}{T_{CU}}\left(2\sum_{t=T_1+1}^{T_2}y_{t-1}\Delta y_t+\rho_b\sum_{t=T_1+1}^{T_2}y_{t-1}^2\right)\CP -\frac{1}{2}(\lambda_r^*-\lambda_1^*)\sigma^2,
\]
which implies that $\mbox{LR}_{a,12}^r\to -\infty$. To investigate $t_{T_1,T_2}$, note that for $T_2 > T_r$,
\begin{align}
    \sum_{t=T_1+1}^{T_2}y_{t-1}^2
    & = 
    \sum_{t=T_1+1}^{T_r}y_{t-1}^2+\sum_{t=T_r+1}^{T_2}y_{t-1}^2 \nonumber \\
    & \sim_a
    T_{CU}T^{\beta}(\lambda_r^*-\lambda_1^*)\frac{\sigma^2}{2b}+T_{CU}^2\sigma^2\int_{\lambda_r^*}^{\lambda_2^*}(W(s^*)-W(\lambda_r^*))^2ds^*,
    \label{app:recovery:sum_y2:T1T2:H1:ba}
\end{align}
while
\begin{align}
    \sum_{t=T_1+1}^{T_2}y_{t-1}\Delta y_t
    & = 
    \sum_{t=T_1+1}^{T_r}y_{t-1}\Delta y_t+\sum_{t=T_r+1}^{T_2}y_{t-1}\Delta y_t \nonumber \\
    & \sim_a
    -T_{CU}\frac{\sigma^2}{2}(\lambda_r^*-\lambda_1^*)+T_{CU}\frac{\sigma^2}{2}\left[\left(W(\lambda_2^*)-W(\lambda_r^*)\right)^2-(\lambda_2^*-\lambda_r^*)\right],
    \label{app:recovery:sum_ydy:T1T2:H1:ba}
\end{align}
which implies $t_{T_1,T_2}=O_p(1)$ for $T_2 > T_r$. On the contrary, for $T_1 < T_2 \leq T_r$, only the first terms of \eqref{app:recovery:sum_y2:T1T2:H1:ba} and \eqref{app:recovery:sum_ydy:T1T2:H1:ba} matter and we can observe that $t_{T_1,T_2}\to -\infty$. Therefore, we conclude that $\mbox{EM}_{a,12},\;\mbox{EM}_{b,12}\to -\infty$. The behavior of $\mbox{LR}_{a,21}$, $\mbox{EM}_{a,21}$, and $\mbox{EM}_{b,21}$ is obtained similarly to the proof of Theorem \ref{theorem:recovery:H0}(ii) and thus they diverge to $-\infty$.

\noindent
(ii-b) Suppose that $T_r < T_1$ ($\lambda_r < \lambda_1)$. Then, following the proof of Theorem \ref{theorem:recovery:H0}(ii), we observe that $\mbox{LR}_{a,12}\to\infty$ and $\mbox{EM}_{a,12}$, and $\mbox{EM}_{b,12}$ are $O_p(1)$.

For $T_2 < T_r$, we observe that
\begin{align}
    \sum_{t=T_2+1}^{T_1}y_{t-1}^2
    & =
    \sum_{t=T_2+1}^{T_r}y_{t-1}^2+\sum_{t=T_r+1}^{T_1}y_{t-1}^2 \nonumber \\
    & \sim_a
    T_{CU}T^{\beta}(\lambda_r^*-\lambda_2^*)\frac{\sigma^2}{2b}+T_{CU}^2\int_{\lambda_r^*}^{\lambda_1^*}\left(W(s^*)-W(\lambda_r^*)\right)^2ds^*
    \label{app:recovery:sum_y2:T2T1:H1:ba}
\end{align}
and
\begin{align}
    \sum_{t=T_2+1}^{T_1}y_{t-1}\Delta y_t
    & = 
    \sum_{t=T_2+1}^{T_r}y_{t-1}\Delta y_t+\sum_{t=T_r+1}^{T_1}y_{t-1}\Delta y_t \nonumber \\
    & \sim_a
    -T_{CU}\frac{\sigma^2}{2}(\lambda_r^*-\lambda_2^*)+T_{CU}\frac{\sigma^2}{2}\left[\left(W(\lambda_1^*)-W(\lambda_r^*)\right)^2-(\lambda_1^*-\lambda_r^*)\right].
    \label{app:recovery:sum_ydy:T2T1:H1:ba}
\end{align}
Then, we observe that $\mbox{LR}_{a,21} \to \infty$. On the contrary, $t_{T_2,T_1}=O_p(1)$ for $T_2 < T_r$ and the same order is observed for $T_r\leq T_2$. As the sign of $t_{T_2,T_1}$ is indeterministic while the scalars go to zero from the definitions of $\mbox{EM}_{a,21}^r$ and $\mbox{EM}_{b,21}^r$, we conclude that they diverge to $\infty$ or $-\infty$.$\blacksquare$

\newpage
\begin{table}[t]
    \centering
        \caption{Coefficients of the Response Surface Regressions}
        \medskip
        \begin{tabular}{crrrrr}
            \hline\hline
            & $a_{0,\ell}$ & $a_{-1,\ell}$ & $a_{1,\ell}$ & $a_{2,\ell}$ & $a_{3,\ell}$ \\
            $\mbox{LR}_{a,21}^e$ & $-9.99\times 10^{-4}$ & $5.13\times 10^{-5}$ & $-1.09\times 10^{-3}$ & $4.40\times 10^{-4}$ & $-2.16\times 10^{-4}$ \\
            $\mbox{EM}_{a,21}^e$ & $-0.127$ & $-4.75\times 10^{-4}$ & $1.34$ & $-0.185$ & $0.0956$ \\
            $\mbox{EM}_{b,21}^e$ & $1.59$ & $-0.0368$ & $0.706$ & $-0.525$ & $0.194$ \\
            \\
            $\mbox{EM}_{a,12}^r$ & $-1.47$ & $5.02\times 10^{-5}$ & $1.57$ & $-0.0124$ & $0.0779$ \\
            \\
            $\mbox{EM}_{b,12}^r$ & $-2.81$ & $-7.44\times 10^{-5}$ & $0.258$ & $-0.382$ & $0.745$ \\
            & $b_{0,\ell}$ & $b_{-1,\ell}$ & $b_{1,\ell}$ & $b_{2,\ell}$ & $b_{3,\ell}$ \\
        &  $-2710$ & $530$ & $5192$ & $-4420$ & $1411$ \\
        \hline     
        \end{tabular}
    \label{table:response}
\end{table}
\hspace{500mm}

\afterpage{\clearpage}
\newpage
\begin{table}[ht]
\begin{center}
\caption{Coverage Rates and Lengths of the Confidence Sets ($T_e$, Case 1, true ends)}
\medskip
\label{table:Te:case1:true}
\begin{tabular}{lccccc}
\hline
           &   $a$ & $\mbox{LR}_a^e$ & $\mbox{EM}_a^e$ & $\mbox{EM}_b^e$ & $\mbox{LE}^e$ \\
\hline
coverage     &  2  & 0.19  &  0.70  &  0.45  &  0.90\\
coverage12   &     & 0.39  &  0.76  &  0.52  &  0.96\\
coverage21   &     & 0.56  &  0.94  &  0.90  &  0.94\\
\hline
coverage     &  4  & 0.33  &  0.86  &  0.68  &  0.91\\
coverage12   &     & 0.55  &  0.92  &  0.77  &  0.97\\
coverage21   &     & 0.67  &  0.94  &  0.90  &  0.94\\
\hline
coverage     &  6  & 0.56  &  0.90  &  0.79  &  0.92\\
coverage12   &     & 0.71  &  0.96  &  0.89  &  0.98\\
coverage21   &     & 0.81  &  0.94  &  0.90  &  0.94\\
\hline
\hline
length         &  2  & 0.09  &  0.42  &  0.21  &  0.61\\
length12left   &     & 0.09  &  0.36  &  0.13  &  0.48\\
length12right  &     & 0.15  &  0.20  &  0.19  &  0.31\\
length21left   &     & 0.31  &  0.47  &  0.45  &  0.47\\
length21right  &     & 0.15  &  0.29  &  0.27  &  0.29\\
\hline
length         &  4  & 0.07  &  0.50  &  0.22  &  0.50\\
length12left   &     & 0.05  &  0.45  &  0.15  &  0.39\\
length12right  &     & 0.22  &  0.27  &  0.27  &  0.31\\
length21left   &     & 0.35  &  0.47  &  0.45  &  0.47\\
length21right  &     & 0.09  &  0.20  &  0.17  &  0.20\\
\hline
length         &  6  & 0.08  &  0.51  &  0.21  &  0.39\\
length12left   &     & 0.05  &  0.51  &  0.16  &  0.29\\
length12right  &     & 0.27  &  0.30  &  0.29  &  0.31\\
length21left   &     & 0.41  &  0.47  &  0.45  &  0.47\\
length21right  &     & 0.07  &  0.14  &  0.11  &  0.14\\
\hline
\end{tabular}
\end{center}
Note: $\mbox{LE}^e$ is a combination of $\mbox{LR}_{b,12}^e$ and $\mbox{EM}_{a,21}^e$.
\end{table}

\afterpage{\clearpage}
\newpage
\begin{table}[ht]
\begin{center}
\caption{Coverage Rates and Lengths of the Confidence Sets ($T_e$, Case 1, estimated ends)}
\medskip
\label{table:Te:case1:est}
\begin{tabular}{lccccc}
\hline
           &   $a$ & $\mbox{LR}_a^e$ & $\mbox{EM}_a^e$ & $\mbox{EM}_b^e$ & $\mbox{LE}^e$ \\
\hline
coverage     &  2  & 0.21 &   0.58  &  0.35  &  0.71\\
coverage12   &     & 0.36 &   0.64  &  0.42  &  0.77\\
coverage21   &     & 0.50 &   0.77  &  0.74  &  0.77\\
\hline               
coverage     &  4  & 0.34 &   0.81  &  0.63  &  0.86\\
coverage12   &     & 0.54 &   0.87  &  0.73  &  0.92\\
coverage21   &     & 0.63 &   0.88  &  0.84  &  0.88\\
\hline               
coverage     &  6  & 0.54 &   0.89  &  0.78  &  0.91\\
coverage12   &     & 0.71 &   0.94  &  0.87  &  0.96\\
coverage21   &     & 0.77 &   0.92  &  0.89  &  0.92\\
\hline
\hline
length         &  2  & 0.12 &   0.42  &  0.19  &  0.60\\
length12left   &     & 0.10 &   0.38  &  0.13  &  0.46\\
length12right  &     & 0.15 &   0.19  &  0.16  &  0.32\\
length21left   &     & 0.29 &   0.44  &  0.42  &  0.44\\
length21right  &     & 0.19 &   0.32  &  0.30  &  0.32\\
\hline                 
length         &  4  & 0.09 &   0.50  &  0.22  &  0.52\\
length12left   &     & 0.06 &   0.47  &  0.16  &  0.41\\
length12right  &     & 0.21 &   0.25  &  0.24  &  0.30\\
length21left   &     & 0.35 &   0.47  &  0.46  &  0.47\\
length21right  &     & 0.11 &   0.21  &  0.17  &  0.21\\
\hline                 
length         &  6  & 0.08 &   0.52  &  0.21  &  0.41\\
length12left   &     & 0.05 &   0.52  &  0.16  &  0.31\\
length12right  &     & 0.25 &   0.27  &  0.27  &  0.29\\
length21left   &     & 0.43 &   0.49  &  0.47  &  0.49\\
length21right  &     & 0.07 &   0.14  &  0.11  &  0.14\\
\hline
\end{tabular}
\end{center}
Note: $\mbox{LE}^e$ is a combination of $\mbox{LR}_{b,12}^e$ and $\mbox{EM}_{a,21}^e$.
\end{table}

\afterpage{\clearpage}
\newpage
\begin{table}[ht]
\begin{center}
\caption{Coverage Rates and Lengths of the Confidence Sets ($T_c$, Case 1, true ends)}
\medskip
\label{table:Tc:case1:true}
\begin{tabular}{lccccc}
\hline
           &   $a$ & $\mbox{LR}_a^c$ & $\mbox{EM}_a^c$ & $\mbox{EM}_b^c$ & $\mbox{LE}^c$ \\
\hline
coverage     &  2  & 0.38  &  0.76  &  0.36 & 0.61  \\
coverage12   &     & 0.59  &  0.91  &  0.58 & 0.91  \\
coverage21   &     & 0.65  &  0.81  &  0.57 & 0.65  \\
\hline                                      
coverage     &  4  & 0.82  &  0.95  &  0.82 & 0.91  \\
coverage12   &     & 0.87  &  0.99  &  0.90 & 0.99  \\
coverage21   &     & 0.92  &  0.96  &  0.88 & 0.92  \\
\hline                                      
coverage     &  6  & 0.95  &  0.98  &  0.96 & 0.97  \\
coverage12   &     & 0.96  &  1.00  &  0.97 & 1.00  \\
coverage21   &     & 0.98  &  0.98  &  0.97 & 0.98  \\
\hline
\hline
length         &  2  & 0.07  &  0.29  &  0.05 & 0.20  \\
length12left   &     & 0.08  &  0.23  &  0.06 & 0.23  \\
length12right  &     & 0.27  &  0.36  &  0.27 & 0.36  \\
length21left   &     & 0.26  &  0.30  &  0.23 & 0.26  \\
length21right  &     & 0.10  &  0.16  &  0.04 & 0.10  \\
\hline                                        
length         &  4  & 0.08  &  0.30  &  0.07 & 0.22  \\
length12left   &     & 0.05  &  0.19  &  0.05 & 0.19  \\
length12right  &     & 0.37  &  0.41  &  0.37 & 0.41  \\
length21left   &     & 0.35  &  0.37  &  0.33 & 0.35  \\
length21right  &     & 0.07  &  0.15  &  0.05 & 0.07  \\
\hline                                        
length         &  6  & 0.09  &  0.28  &  0.08 & 0.20  \\
length12left   &     & 0.05  &  0.16  &  0.05 & 0.16  \\
length12right  &     & 0.41  &  0.42  &  0.41 & 0.42  \\
length21left   &     & 0.39  &  0.39  &  0.38 & 0.39  \\
length21right  &     & 0.06  &  0.14  &  0.05 & 0.06  \\
\hline
\end{tabular}
\end{center}
\end{table}

\afterpage{\clearpage}
\newpage
\begin{table}[ht]
\begin{center}
\caption{Coverage Rates and Lengths of the Confidence Sets ($T_c$, Case 1, estimated ends)}
\medskip
\label{table:Tc:case1:est}
\begin{tabular}{lccccc}
\hline
           &   $a$ & $\mbox{LR}_a^c$ & $\mbox{EM}_a^c$ & $\mbox{EM}_b^c$ & $\mbox{LE}^c$ \\
\hline
coverage     &  2  & 0.32  &  0.55  &  0.32 & 0.46 \\
coverage12   &     & 0.46  &  0.66  &  0.47 & 0.66 \\
coverage21   &     & 0.50  &  0.59  &  0.46 & 0.50 \\
\hline                                     
coverage     &  4  & 0.78  &  0.85  &  0.78 & 0.84 \\
coverage12   &     & 0.80  &  0.89  &  0.83 & 0.89 \\
coverage21   &     & 0.84  &  0.86  &  0.82 & 0.84 \\
\hline                                     
coverage     &  6  & 0.92  &  0.94  &  0.93 & 0.94 \\
coverage12   &     & 0.93  &  0.96  &  0.94 & 0.96 \\
coverage21   &     & 0.94  &  0.94  &  0.94 & 0.94 \\
\hline
\hline
length         &  2  & 0.07  &  0.25  &  0.06 & 0.18 \\
length12left   &     & 0.08  &  0.21  &  0.07 & 0.21 \\
length12right  &     & 0.30  &  0.37  &  0.28 & 0.37 \\
length21left   &     & 0.28  &  0.31  &  0.24 & 0.28 \\
length21right  &     & 0.10  &  0.16  &  0.05 & 0.10 \\
\hline                                       
length         &  4  & 0.09  &  0.28  &  0.08 & 0.20 \\
length12left   &     & 0.05  &  0.18  &  0.05 & 0.18 \\
length12right  &     & 0.39  &  0.42  &  0.38 & 0.42 \\
length21left   &     & 0.36  &  0.36  &  0.34 & 0.36 \\
length21right  &     & 0.07  &  0.15  &  0.05 & 0.07 \\
\hline                                       
length         &  6  & 0.09  &  0.28  &  0.09 & 0.20 \\
length12left   &     & 0.05  &  0.16  &  0.05 & 0.16 \\
length12right  &     & 0.40  &  0.42  &  0.40 & 0.42 \\
length21left   &     & 0.39  &  0.39  &  0.38 & 0.39 \\
length21right  &     & 0.06  &  0.14  &  0.06 & 0.06 \\
\hline
\end{tabular}
\end{center}
\end{table}

\afterpage{\clearpage}
\newpage
\begin{table}[ht]
\begin{center}
\caption{Coverage Rates and Lengths of the Confidence Sets ($T_r$, Case 1, true ends)}
\medskip
\label{table:Tr:case1:true}
\begin{tabular}{lccccc}
\hline
           &   $a$ & $\mbox{LR}_a^r$ & $\mbox{EM}_a^r$ & $\mbox{EM}_b^r$ & $\mbox{LE}^r$ \\
\hline
coverage     &  2  & 0.20  &  0.76  &  0.63  &  0.93\\
coverage12   &     & 0.45  &  0.93  &  0.97  &  0.97\\
coverage21   &     & 0.49  &  0.82  &  0.65  &  0.95\\
\hline
coverage     &  4  & 0.31  &  0.89  &  0.84  &  0.94\\
coverage12   &     & 0.50  &  0.92  &  0.96  &  0.96\\
coverage21   &     & 0.66  &  0.97  &  0.87  &  0.97\\
\hline
coverage     &  6  & 0.49  &  0.91  &  0.92  &  0.95\\
coverage12   &     & 0.61  &  0.92  &  0.96  &  0.96\\
coverage21   &     & 0.81  &  0.99  &  0.96  &  0.98\\
\hline
\hline
length         &  2  & 0.09  &  0.45  &  0.30  &  0.62\\
length12left   &     & 0.11  &  0.30  &  0.34  &  0.34\\
length12right  &     & 0.28  &  0.48  &  0.52  &  0.52\\
length21left   &     & 0.12  &  0.14  &  0.16  &  0.22\\
length21right  &     & 0.13  &  0.43  &  0.19  &  0.50\\
\hline
length         &  4  & 0.06  &  0.53  &  0.33  &  0.55\\
length12left   &     & 0.06  &  0.18  &  0.21  &  0.21\\
length12right  &     & 0.31  &  0.48  &  0.52  &  0.52\\
length21left   &     & 0.17  &  0.20  &  0.20  &  0.22\\
length21right  &     & 0.10  &  0.51  &  0.22  &  0.44\\
\hline
length         &  6  & 0.07  &  0.54  &  0.31  &  0.44\\
length12left   &     & 0.04  &  0.12  &  0.13  &  0.13\\
length12right  &     & 0.35  &  0.48  &  0.52  &  0.52\\
length21left   &     & 0.20  &  0.21  &  0.22  &  0.22\\
length21right  &     & 0.09  &  0.56  &  0.23  &  0.35\\
\hline
\end{tabular}
\end{center}
Note: $\mbox{LE}^r$ is a combination of $\mbox{EM}_{b,12}^r$ and $\mbox{LR}_{b,21}^r$.
\end{table}

\afterpage{\clearpage}
\newpage
\begin{table}[ht]
\begin{center}
\caption{Coverage Rates and Lengths of the Confidence Sets ($T_r$, Case 1, estimated ends)}
\medskip
\label{table:Tr:case1:est}
\begin{tabular}{lccccc}
\hline
           &   $a$ & $\mbox{LR}_a^r$ & $\mbox{EM}_a^r$ & $\mbox{EM}_b^r$ & $\mbox{LE}^r$ \\
\hline
coverage     &  2  & 0.20  &  0.63  &  0.51  &  0.71\\
coverage12   &     & 0.39  &  0.76  &  0.79  &  0.79\\
coverage21   &     & 0.40  &  0.67  &  0.53  &  0.73\\
\hline               
coverage     &  4  & 0.33  &  0.83  &  0.79  &  0.87\\
coverage12   &     & 0.50  &  0.86  &  0.91  &  0.91\\
coverage21   &     & 0.63  &  0.90  &  0.82  &  0.90\\
\hline               
coverage     &  6  & 0.50  &  0.89  &  0.90  &  0.92\\
coverage12   &     & 0.61  &  0.89  &  0.94  &  0.94\\
coverage21   &     & 0.79  &  0.97  &  0.93  &  0.96\\
\hline
\hline
length         &  2  & 0.10  &  0.48  &  0.30  &  0.61\\
length12left   &     & 0.14  &  0.31  &  0.35  &  0.35\\
length12right  &     & 0.26  &  0.47  &  0.50  &  0.50\\
length21left   &     & 0.12  &  0.14  &  0.13  &  0.21\\
length21right  &     & 0.13  &  0.45  &  0.20  &  0.49\\
\hline                 
length         &  4  & 0.08  &  0.53  &  0.32  &  0.55\\
length12left   &     & 0.07  &  0.19  &  0.21  &  0.21\\
length12right  &     & 0.32  &  0.50  &  0.53  &  0.53\\
length21left   &     & 0.14  &  0.16  &  0.16  &  0.19\\
length21right  &     & 0.10  &  0.53  &  0.23  &  0.46\\
\hline                 
length         &  6  & 0.08  &  0.55  &  0.31  &  0.45\\
length12left   &     & 0.05  &  0.12  &  0.13  &  0.13\\
length12right  &     & 0.37  &  0.51  &  0.54  &  0.54\\
length21left   &     & 0.16  &  0.17  &  0.18  &  0.18\\
length21right  &     & 0.09  &  0.57  &  0.24  &  0.36\\
\hline
\end{tabular}
\end{center}
Note: $\mbox{LE}^r$ is a combination of $\mbox{EM}_{b,12}^r$ and $\mbox{LR}_{b,21}^r$.
\end{table}

\afterpage{\clearpage}
\newpage

\afterpage{\clearpage}
\newpage

\afterpage{\clearpage}
\newpage

\end{document}